\documentclass[preprint2]{aastex}
\usepackage{multirow}
\usepackage{subfigure, apalike}
\usepackage{amssymb}
\usepackage{natbib}
\usepackage{multirow}
\usepackage{geometry}
\usepackage{amsmath}
\usepackage{color}

\begin{document}


\title{On Presolar Stardust Grains from CO Classical Novae}


\author{Christian~Iliadis}
\affil{Department of Physics \& Astronomy, University of North Carolina, Chapel Hill, NC 27599-3255}
\affil{Triangle Universities Nuclear Laboratory, Durham, NC 27708-0308, USA}
\email{iliadis@unc.edu}

\author{Lori N.~Downen}
\affil{Department of Physics \& Astronomy, University of North Carolina, Chapel Hill, NC 27599-3255}
\affil{Triangle Universities Nuclear Laboratory, Durham, NC 27708-0308, USA}

\author{Jordi~Jos\'{e}}
\affil{Departament de F\'\i sica, EEBE, Universitat Polit\`ecnica de Catalunya, c/Eduard Maristany 10,
E-08930 Barcelona,
Spain}
\affil{Institut d'Estudis Espacials de Catalunya, c/Gran Capit\`a 2-4, Ed. Nexus-201, E-08034 Barcelona, Spain}

\author{Larry R.~Nittler}
\affil{Department of Terrestrial Magnetism, Carnegie Institution for Science, Washington, DC 20015, USA}

\author{Sumner~Starrfield}
\affil{Earth and Space Exploration, Arizona State University, Tempe, AZ 85287-1404, USA}




\begin{abstract}
About 30\% to 40\% of classical novae produce dust $20$ $-$ $100$ days after the outburst, but no presolar stardust grains from classical novae have been unambiguously identified yet. Although several studies claimed a nova paternity for certain grains, the measured and simulated isotopic ratios could only be reconciled assuming that the grains condensed after the nova ejecta mixed with a much larger amount of close-to-solar matter. However, the source and mechanism of this potential post-explosion dilution of the ejecta remains a mystery. A major problem with previous studies is the small number of simulations performed and the implied poor exploration of the large nova parameter space. We report the results of a different strategy, based on a Monte Carlo technique, that involves the random sampling over the most important nova model parameters: the white dwarf composition; the mixing of the outer white dwarf layers with the accreted material before the explosion; the peak temperature and density; the explosion time scales; and the possible dilution of the ejecta after the outburst. We discuss and take into account the systematic uncertainties for both the presolar grain measurements and the simulation results. Only those simulations that are consistent with all measured isotopic ratios of a given grain are accepted for further analysis. We also present the numerical results of the model parameters. We identify $18$ presolar grains with measured isotopic signatures consistent with a CO nova origin, {\it without assuming any dilution of the ejecta}. Among these, the grains G270$\_$2, M11-334-2, G278, M11-347-4, M11-151-4, and Ag2$\_$6 have the highest probability of a CO nova paternity.
\end{abstract}




\section{Introduction}\label{sec:intro}
Primitive meteorites contain dust grains with isotopic compositions vastly different from those of any other matter found in the solar system \citep{zinner14}. The only viable explanation for their existence is that they condensed in stellar winds or the ejecta of exploding stars. These tiny grains survived the journey through the interstellar medium to the region in which the presolar cloud formed about $4.6$~Gy ago. Some of these grains also survived the homogenization process during the formation of the solar system and were incorporated into meteorites. They are called presolar stardust grains and retain the distinct isotopic composition of the stellar outflows at the time of grain condensation. The laboratory measurement of their isotopic ratios provides an exceptional opportunity to study stellar evolution, stellar explosions, nucleosynthesis, dust formation, and galactic chemical evolution. 

The analysis and interpretation of presolar stardust grains requires an iterative approach \citep{nittler16}. First, the stellar source for a group of grains needs to be identified on the basis of the available isotopic data. Once the source is identified, the precisely measured isotopic ratios provide strong constraints for understanding the physical and chemical processes that occurred inside the parent stars. According to current thinking, most analyzed stardust grains formed in Asymptotic Giant Branch (AGB) stars of a previous generation. This insight was important for a quantitative understanding of AGB stars, and also for demonstrating how one half of all elements beyond iron are synthesized in the astrophysical s-process. A fraction of the measured stardust grains ($\approx$ $1$\% $-$ $5$\% of SiC; up to 10\% of oxides and silicates; $\approx$ $50$\% of graphite) originate presumably from core-collapse supernovae \citep{zinner14}. Their isotopic signatures may shed light on explosive nucleosynthesis, the mixing between different layers in the ejecta, grain condensation, and how much dust survived the reverse shock before injection into the interstellar medium.  

A few presolar stardust grains are characterized by very low $^{12}$C/$^{13}$C and $^{14}$N/$^{15}$N ratios and large $^{30}$Si excesses \citep{amari01}. These signatures imply an increased production of the minor isotopes, $^{13}$C and $^{15}$N, compared to the major ones, $^{12}$C and $^{14}$N, and are difficult to explain by AGB star or supernova nucleosynthesis. Explosive hydrogen burning in classical novae, on the other hand, seems to reproduce qualitatively some of the isotopic signatures measured in these grains \citep{amari01,jose04,jose07,haenecour16}.

\begin{figure}
\epsscale{1.0}
\plotone{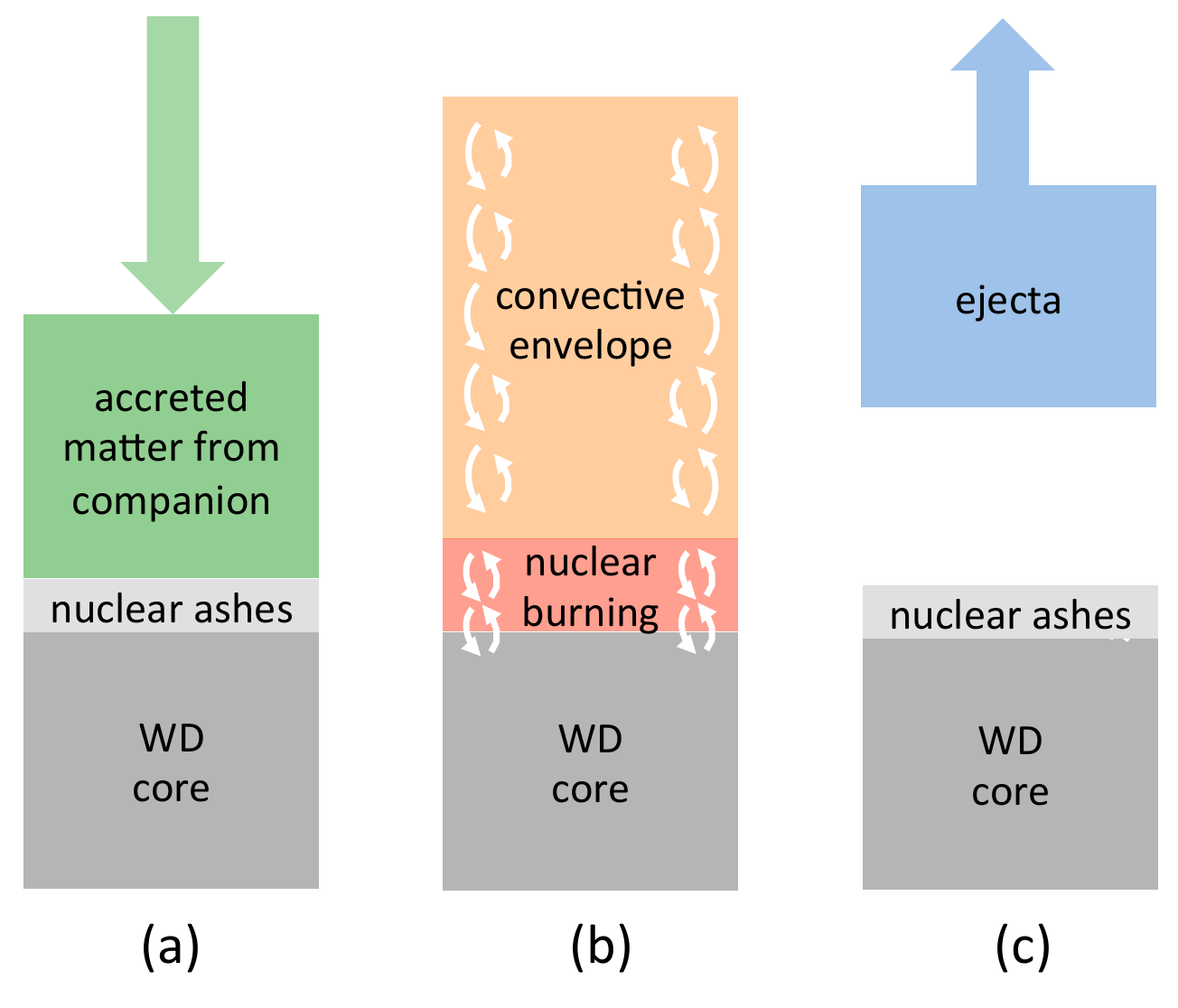}
\caption{Sketch of processes during a classical nova outburst. 
(a) Nuclear ashes from a previous outburst (He plus metals; light grey region) sit on top of a CO-rich white dwarf core (dark grey region), consisting mainly of $^{12}$C and $^{16}$O; the white dwarf accretes hydrogen-rich matter (green region) from a companion. (b) The temperature increases at the base of the envelope until a thermonuclear runaway (TNR) occurs; during the TNR, mixing and diffusion takes place at the interface of accreted and white dwarf outer core matter (red region); the convective envelope (orange region) extends to the surface. (c) A fraction of the nuclearly processed accreted-plus-core matter is ejected (blue region) and a fraction remains on the white dwarf (light grey region) to take part in the next event.  
 \label{fig:sketch}}
\end{figure}

A classical nova is thought to be one consequence of the accretion of hydrogen-rich material onto a white dwarf in a close binary system \citep{jose16,starrfield16}. Some of the key processes are sketched in Figure~\ref{fig:sketch}. Over long periods of time, the material being accreted from the secondary star forms a layer of nuclear fuel (green region in Figure~\ref{fig:sketch}a) on the white dwarf surface. The bottom of this layer is gradually compressed by the surface gravity and becomes electron degenerate. Once the temperature at the bottom of the accreted layer reaches the Fermi temperature ($\approx$ $30$~MK), the layer begins to expand, but by this time the temperature is increasing so fast that a thermonuclear runaway results. During the steep temperature rise, matter from the outermost white dwarf core layer is dredged up into the accreted matter (red region in Figure~\ref{fig:sketch}b), as first suggested by \citet{ferland78}. This significantly enriches the burning layer in CNO nuclei, which is crucial for ensuring a strong nuclear energy release and a violent outburst; it also helps to explain the observed abundances inferred from the ejecta. The ejected material (blue region in Figure~\ref{fig:sketch}c) consists of a mixture of white dwarf and accreted matter that has been processed by explosive hydrogen burning. 

Spectroscopic studies identified two distinct types of classical novae. Nova ejecta rich in CNO material point to an underlying CO white dwarf, which represents the evolutionary fate of a low-mass star after the cessation of core helium burning (Section~\ref{sec:wd}). These objects are called ``CO novae''. On the other hand, elemental enrichments in the range of Ne to Ar have been attributed to the presence of an underlying, massive ONe white dwarf, representing the evolutionary fate of an intermediate-mass star after completion of core carbon burning. The latter explosions, often referred to as ``neon novae'' or ``ONe novae'', tend to be more energetic than CO novae \citep{starrfield86}.

About $20$-$100$ days after the outburst, many classical nova light curves show a rapid decline in the optical flux because of extinction, and a corresponding rise in the mid-infrared luminosity because of thermal emission \citep{shore12,evans12}. This observation strongly suggests that dust grains, with radii up to $\approx$ $10$~$\mu$m \citep{stratton39,gehrz98,gehrz08}, form in the ejecta when they cool to temperatures below $\approx1700$~K. Overall, about 30-40\% of novae produce dust, including both CO and ONe novae \citep{evans12}. A number of CO novae are known to have been prolific dust producers (see Table~\ref{tab:dust}). When dust forms, its observed mass fraction in nova ejecta is about $10^{-3}$, corresponding to a mass between  $10^{-10}$~M$_\odot$ and $10^{-6}$~M$_\odot$. 

Unlike most other sources, classical novae have been observed to produce carbon-rich and oxygen-rich dust simultaneously \citep{gehrz92}. In principle, the composition of the dust that forms in a given environment depends sensitively on the carbon-to-oxygen ratio. When the number abundance ratio of C/O exceeds unity, and all oxygen atoms are locked up in strongly bound CO molecules, carbon-rich dust forms; similarly, oxygen-rich dust forms when the value of C/O is less than unity \citep{waters04}. This assumes that the CO abundance reaches its saturation value. However, if CO does not form to saturation, neither carbon nor oxygen will be entirely bound in CO molecules, leaving both elements available for dust condensation \citep{evans08}. In addition, \citet{jose04} found that the presence of significant amounts of intermediate-mass elements, such as Al, Ca, Mg, or Si, may dramatically alter the condensation process, allowing for the formation of carbon-rich dust even in an oxygen-rich environment. Table~\ref{tab:dust} summarizes the measured carbon and oxygen abundances (by mass) in CO nova shells, together with the observation of dust species.

A classical nova is not an ideal environment for dust condensation. Diatomic and polyatomic molecules can only form in the ejecta if they are shielded from ionization by the copious UV radiation of the white dwarf, which remains a supersoft X-ray source after the outburst \citep{schwarz11}. The shielding can be provided only by spatially inhomogeneous regions of the ejecta that have a much higher than average gas density. Clumpy ejecta have been inferred spectroscopically for many classical novae \citep{williams92,saizar94}, but the physical origin remains an open question.

The above discussion implies that measuring the isotopic signatures of presolar stardust grains originating from classical novae could shed light on the explosion mechanism, the mixing of matter during the outburst, and the formation of molecules and dust in the expanding ejecta. While several authors have claimed a nova paternity for certain stardust grains, no grains from novae have been unambiguously identified yet. Therefore, they are referred to in the literature as ``nova candidate'' or ``putative nova'' grains. 

A significant problem is that most classical nova simulations result in ejecta with much more anomalous isotopic ratios compared to what has been measured in the grains \citep{nittler05,gyngard10,leitner12,nguyen14}. To explain the measurements, it has been speculated \citep{amari01} that these grains may have condensed after the nova ejecta mixed with a much larger amount ($\gtrsim$ $90$\%) of close-to-solar matter. However, the origin of the latter contribution is not well understood. In addition, counter-arguments favor a supernova origin for some of these ``nova candidate'' grains \citep{nittler05,liu16}. Recently, the ``first plausible grain of CO nova origin'' has been reported, based on the measured C, N, Si, and S isotopic compositions, without requiring any mixing with solar-like matter \citep{haenecour16}. This would imply that dust from classical novae contributed to the building blocks of the solar system. A severe problem with this interpretation is the mismatch of the simulated and measured $^{16}$O/$^{17}$O and  $^{16}$O/$^{18}$O ratios in that particular stardust grain, with the deviations amounting to several orders of magnitude.

Although the total amount of matter ejected by classical novae per year in our galaxy is much smaller when compared to the contributions of AGB stars or type II supernovae, it is puzzling that among several thousand presolar stardust grains identified so far, we cannot claim with confidence a classical nova paternity for a single grain. A major problem is that hydrodynamic nova simulations have a poorly constrained parameter space, and that these CPU-intensive simulations sample a restricted number of parameter combinations before the computed isotopic ratios are compared with stardust grain measurements. Here, we follow a different approach that explores a large region of the nova parameter space. Since we need to perform a large number of simulations, our method does not depend on specific classical nova hydrodynamic modeling, but is by necessity model independent. 

We will focus on CO novae and leave an investigation of ONe novae to future work. In Section~\ref{sec:strategy}, we present our strategy. Section~\ref{sec:procedure} discusses our simulation procedure, together with the various parameters entering the calculations. Results are presented in Section~\ref{sec:results}. A summary and conclusions are given in Section~\ref{sec:summary}. 

\section{Strategy}\label{sec:strategy}
The two questions we attempt to answer are: (1) Does a given presolar stardust grain originate from a classical nova? (2) What are the conditions that gave rise to the measured isotopic ratios? These questions are intricately connected. If we cannot identify any viable nova conditions that could give rise to the data, we may not claim that a given stardust grain has a nova paternity. These questions have been partially addressed using one-dimensional hydrodynamic nova simulations \citep{jose04}. As with any stellar model, such simulations depend on many assumptions and parameters. Some model parameters are constrained by observation or experiment (e.g., thermonuclear reaction rates and the nuclear energy release), while only indirect information is available for other parameters (e.g., the rate of mass accretion from the companion, the initial composition of the fuel, initial luminosity and mass of the white dwarf, the amount of white dwarf matter dredged up into the accreted envelope, and the effects of multicycle nova evolution). Some effects have remained nearly unexplored, e.g., the impact of a magnetic field or rotation on the nova outburst. The simulation of dust formation introduces a host of additional assumptions, e.g., the shielding of molecules from the radiation of the white dwarf, the formation of clumpy ejecta, mixing of the ejecta with matter of the interstellar medium or the accretion disk, and grain nucleation and growth to macroscopic size. The main disadvantage of studying the nova paternity of stardust grains with a hydrodynamic computer code is that only a relatively small number of simulations can be performed. Since the nova parameter space is only sparsely explored, parameter value combinations favorable for reproducing isotopic signatures of nova grains may easily be missed. 

We seek a more comprehensive exploration of the nova parameter space. To this end, our strategy involves three key ingredients: (i) a simple and fast simulation that can be repeated many times using different combinations of parameter values; (ii) the assumption of a reasonable parameter range and the independent sampling of all parameters; and (iii) the comparison of simulated and observed isotopic ratios for {\it all} elements measured in a given stardust grain. The latter point is important: the grains condensed at a given time and location in the expanding ejecta\footnote{We assume in the present work that the presolar grain composition is not significantly altered by ion implantation. While implantation may be important for concentrations of either noble gases \citep{verchovsky03} or trace elements \citep{clayton02}, it is highly unlikely that this process will alter significantly the major-element (e.g., carbon, oxygen, or silicon) isotopic composition of the grains studied here.}. Unless we can explain all the measured isotopic signatures simultaneously, we may not claim a nova paternity. 

In the following, we will discuss each of these ingredients. We start with a description of a schematic model, then add realistic assumptions pertaining to nova outbursts, and finally discuss how to compare our simulation results to stardust grain data.   

\section{Procedures}\label{sec:procedure}
\subsection{Nuclear reaction network and thermonuclear rates}\label{sec:rates}
We compute the nucleosynthesis using a reaction network consisting of 213 nuclides, ranging from p, n, $^4$He, to $^{55}$Cr. These nuclides are linked by $2373$ nuclear interactions (proton and $\alpha$-particle captures, $\beta$-decays, light-particle reactions, etc.). Thermonuclear reaction rates are adopted from STARLIB v6.5 (09/2017)\footnote{The STARLIB site has moved to: \texttt{https://starlib.github.io/Rate-Library/}.}. This library has a tabular format and contains reaction rates and rate probability density functions on a grid of temperatures between $10$~MK and $10$~GK \citep{sallaska13}. The probability densities can be used to derive statistically meaningful reaction rate uncertainties at any desired temperature. Many of the reaction rates important for the present work that are listed in STARLIB have been computed using a Monte Carlo method, which randomly samples all experimental nuclear physics input parameters \citep{longland10}. Most of the reaction rates important for studying hydrogen burning in CO novae are based on experimental nuclear physics information and provide a reliable foundation for robust predictions. For some reactions of interest to classical novae \citep{iliadis15b}, however, experimental rates are not available yet, and the rates included in STARLIB are adopted from nuclear statistical model calculations using the code TALYS \citep{goriely08}. In such cases, a reaction rate uncertainty factor of $10$ is assumed. 

Stellar weak interaction rates, which depend on both temperature and density, for all species in our network are adopted from \citet{oda94} and, if not listed there, from \citet{fuller82}. The stellar weak decay constants are tabulated at temperatures from $T$ = $10$~MK to $30$~GK, and densities of $\rho Y_e$ $=$ $10$ $-$ $10^{11}$~g/cm$^3$, where $Y_e$ denotes the electron mole fraction. For all stellar weak interaction rates, we assumed a factor of two uncertainty. Short-lived nuclides, e.g., $^{13}$N (T$_{1/2}$ $=$ $10$~min), $^{14}$O (T$_{1/2}$ $=$ $71$~s), $^{15}$O (T$_{1/2}$ $=$ $122$~s), $^{17}$F (T$_{1/2}$ $=$ $64$~s), and $^{18}$F (T$_{1/2}$ $=$ $110$~min), present at the end of a network calculation were assumed to decay to their stable daughter nuclides.

To explore the effects of thermonuclear reaction rate uncertainties, we perform some of our calculations by randomly sampling all rates simultaneously using the rate probability densities provided by STARLIB. This method is discussed in detail in \citet{iliadis15} and was recently applied to explain abundance anomalies in globular clusters \citep{iliadis16}. It suffices to mention here that we adopt a lognormal distribution for the nuclear rates, given by
\begin{equation}
f(x) = \frac{1}{\sigma \sqrt{2\pi}} \frac{1}{x} e^{-(\ln x - \mu)^2/(2\sigma^2)}
\end{equation}
where the lognormal parameters $\mu$ and $\sigma$ determine the location and the width, respectively, of the distribution. For a lognormal probability density, samples, $i$, of a nuclear rate, $y$, are computed from  
\begin{equation}
 \label{eq:sample}
  y_{i} = y_{med}  (f.u.)^{p_{i}}
\end{equation}
where $y_{med}$ and $f.u.$ are the median value and the factor uncertainty, respectively, which are both provided by STARLIB. The quantity $p_{i}$ is a random variable that is normally distributed, i.e., according to a Gaussian distribution with an expectation value of zero and a standard deviation of unity. We emphasize that the factor uncertainty of experimental Monte Carlo reaction rates depends explicitly on stellar temperature \citep{iliadis10,longland12}. 

\subsection{A schematic model of explosive hydrogen burning}\label{sec:scheme}
We adopt a simple, one-zone analytical parameterization for the thermodynamic trajectories of the explosion,
\begin{equation}
T(t) = T_{peak}e^{-t/\tau_T}~,~\rho(t) = \rho_{peak}e^{-t/\tau_\rho}
\end{equation}
where $t\geq0$ is the time since peak temperature, T$_{peak}$, or peak density, $\rho_{peak}$; $\tau_T$ and $\tau_\rho$ are the times at which temperature and density, respectively, have fallen to $1/e$ of their peak values. Notice that we do not assume an adiabatic expansion since we treat $\tau_T$ and $\tau_\rho$ as independent parameters. This is consistent with the results of one-dimensional hydrodynamic nova simulations, which predict non-adiabatic $T$ $-$ $\rho$ evolutions.

It is necessary to demonstrate that our simple simulation has some predictive power as regards to nova nucleosynthesis. In a first step, we generated a hydrodynamic CO nova model using the one-dimensional code SHIVA \citep{jose98}, assuming a white dwarf mass and initial luminosity of $M_{WD}$ $=$ $1.0$~$M_\odot$ and $L_{WD}$ $=$ $10^{-2}$~$L_\odot$, respectively, and accretion of solar-like material at a rate of $\dot{M}_{acc}$ $=$ $2\times10^{-10}$~$M_\odot$ yr$^{-1}$. The composition of the nuclear fuel was obtained by pre-mixing equal amounts of solar-like matter with carbon-oxygen white dwarf matter (assumed to be 50\% $^{12}$C and 50\% $^{16}$O, by mass). The model included 45 envelope zones containing all material involved in the thermonuclear runaway. The deepest zone achieved a peak temperature of $179$~MK, while the innermost ejected zone reached a peak temperature of $163$~MK. Final isotopic abundances for matter that exceeds escape velocity (i.e., the fraction of the envelope ejected) are determined $1$~hr after peak temperature is achieved, and the abundance of each nuclide is mass-averaged over all ejected zones. 

\begin{figure*}
\epsscale{2.0}
\plotone{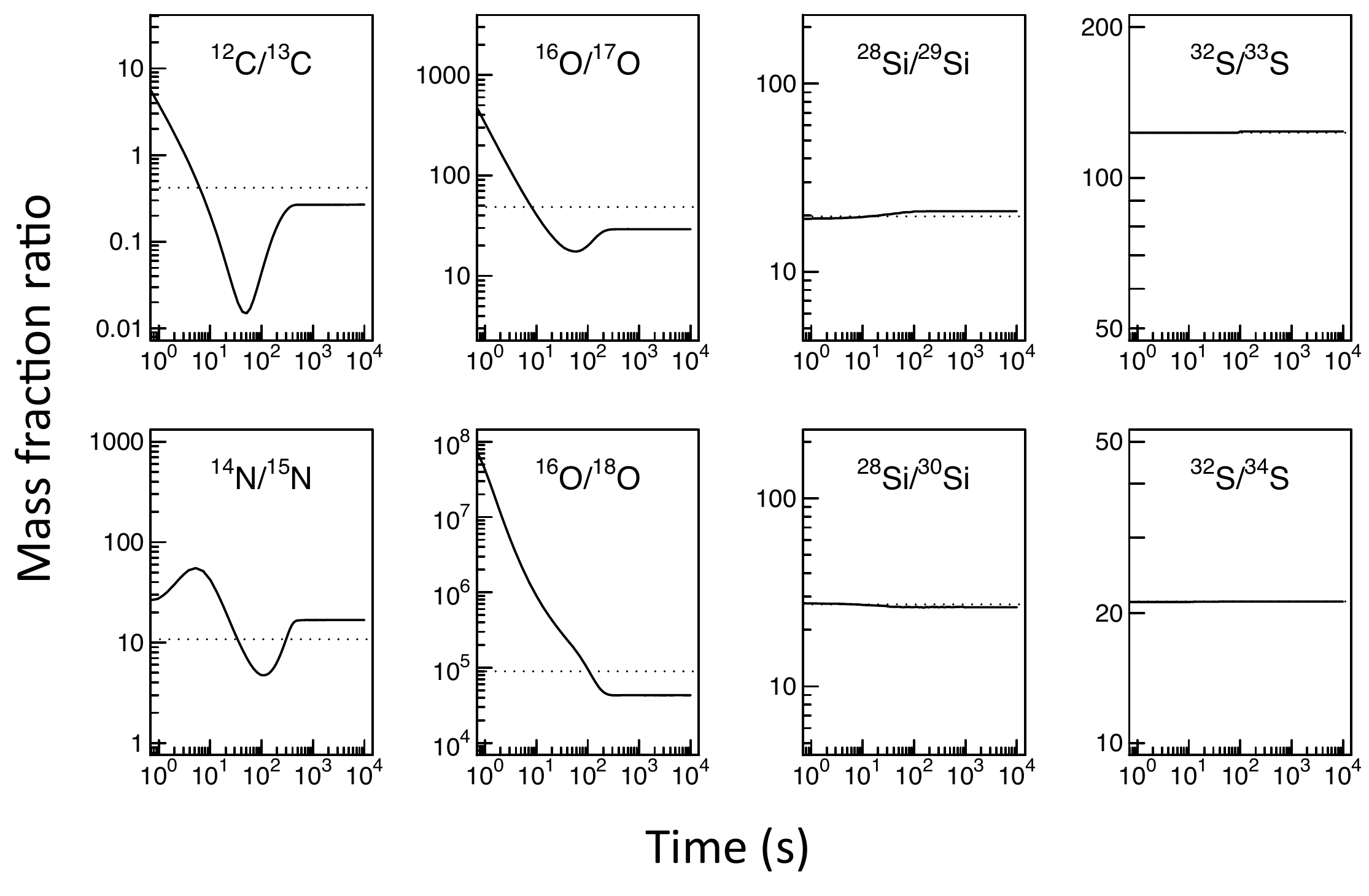}
\caption{Time evolution of C, N, O, Si, and S isotopic ratios as predicted by a parametric, one-zone simulation (solid lines), assuming exponentially decaying temperature and density trajectories. The simulation parameters were T$_{peak}$ $=$ $177$~MK, $\rho_{peak}$ $=$ $200$~g/cm$^3$, $\tau_T$ $=$ $2500$~s, $\tau_\rho$ $=$ $38$~s. Initial abundances were obtained by pre-mixing solar-like matter with carbon-oxygen white dwarf matter (assumed to be 50\% $^{12}$C and 50\% $^{16}$O, by mass) in equal amounts. The dotted lines show the final, mass-zone-averaged results of a one-dimensional hydrodynamic simulation using the same initial composition, where the deepest zone reached a peak temperature of $179$~MK during the outburst. 
\label{fig:test}}
\end{figure*}

In a second step, we adjusted the parameters of our simple one-zone simulation by trial and error to see if we can approximately reproduce the final isotopic ratios of the multi-zone hydrodynamic model, assuming exactly the same initial abundances in both calculations. The resulting isotopic ratios for the most important elements (C, N, O, Si, and S) are shown in Figure~\ref{fig:test}. The solid lines correspond to the time evolutions predicted by the simple one-zone simulation and were obtained with the following parameter values: T$_{peak}$ $=$ $177$~MK, $\rho_{peak}$ $=$ $200$~g/cm$^3$, $\tau_T$ $=$ $2500$~s, $\tau_\rho$ $=$ $38$~s. The total time was 10,000 s, but the results are not sensitive to this parameter once peak temperature and density have significantly declined from their peak values. The dotted line in each panel indicates the mass-zone-averaged ejected final abundance ratios predicted by the multi-zone hydrodynamic calculation. The interesting finding is that the one-zone  simulation reproduces the results of the multi-zone hydrodynamic calculation {\it within a factor of 2}. We repeated the test for other CO white dwarf masses, and even for models of ONe novae, and again obtained agreement within a factor of $2$.

This level of agreement may be at first surprising, considering that our simulation, unlike the hydrodynamic model, follows a single zone only, and disregards accretion, convection, and ejection of matter. However, recall that we are mainly interested in isotopic {\it ratios} instead of absolute isotopic or elemental abundances, which are very likely more sensitive to such effects. We emphasize that the parameters (T$_{peak}$, $\rho_{peak}$, $\tau_T$, $\tau_\rho$) derived from the simple simulation correspond neither to the averages over different mass zones, nor to a given zone, of the hydrodynamic simulation. They nevertheless provide, albeit crude, approximations of the physical conditions during the nuclear burning, mainly because thermonuclear reaction rates are highly sensitive to the plasma temperature. 

It is also interesting to note that the value $\rho_{peak}$ $=$ $200$~g/cm$^3$ of the exponentially decaying density profile in Figure~\ref{fig:test} does not correspond to any maximum density in the one-dimensional hydrodynamic simulation, but is approximately equal to the {\it density at maximum temperature} in the innermost zones of the hydrodynamic calculation. In the latter simulation, the density declines from a maximum value, which typically amounts to a few thousand grams per cubic centimeter and is achieved well before the temperature peaks, to a very small value.

A factor of two agreement is sufficient for the purposes of the present work, as will be shown below. The advantage of our simple procedure is that we can independently sample over the parameters and repeat the simulation many times. If a given stardust grain has indeed a nova paternity, we would expect that certain combinations of parameter values (T$_{peak}$, $\rho_{peak}$, $\tau_T$, $\tau_\rho$) approximately reproduce the measured isotopic ratios, with magnitudes near the ranges typical for classical novae. 

Before we can discuss the presolar stardust grain data, however, we need to introduce three more parameters that will be important for our study: the $^{12}$C/$^{16}$O ratio of the outer white dwarf core, the mixing of matter at the interface of the white dwarf and the envelope, and the dilution of the ejecta by mixing with solar-like matter. 

\subsection{Key parameters for nova nucleosynthesis}
\subsubsection{The white dwarf composition}\label{sec:wd}
Stars with masses between $\approx$ $0.8$ $-$ $8$~M$_\odot$ undergo hydrogen and helium burning in their cores and end their lives as white dwarfs \citep{karakas14}, composed of carbon and oxygen (CO white dwarfs). The composition of the white dwarf depends sensitively on the $^{12}$C($\alpha$,$\gamma$)$^{16}$O reaction rate. Most CO nova studies assume a white dwarf core composition of 50\% $^{12}$C and 50\% $^{16}$O, by mass. A rare exception is the work of \citet{kovetz97}, who performed nova simulations for core compositions of pure $^{12}$C, pure $^{16}$O, and an equal mixture of $^{12}$C and $^{16}$O. However, what is most relevant for nova simulations is the composition of the {\it outermost core} of the white dwarf, since only this layer is expected to be dredged up during the outburst. Recently, \citet{jose16} evolved an $8$~M$_\odot$ progenitor star through successive hydrogen burning, helium burning, and thermally pulsing asymptotic giant branch phases, and used the resulting outer core composition at several locations of the nascent white dwarf as starting points of the CO nova simulations. This resulted in carbon-rich ejecta and the possibility of the formation of carbon-rich dust.

One problem with this assumption is that the white dwarf needs some time to cool before a nova outburst can take place; if the white dwarf is initially too luminous, the envelope is not highly degenerate and only a mild thermonuclear runaway with no mass ejection will occur. For this reason, almost all nova simulations have been performed with an initial white dwarf luminosity in the range of $L_{WD}$ $=$ $10^{-3}$ $-$ $10^{-2}$~$L_\odot$. A few studies assumed higher luminosities, see \citet{starrfield85,yaron05,hernanz08}. The important point is that the composition of the outer core changes while the white dwarf evolves on its cooling track. This question was studied by \citet{bravo11} in connection with models for thermonuclear supernovae. The outer core composition of their $1$~M$_\odot$ model white dwarf changed from a $^{12}$C/$^{16}$O mass fraction ratio of $1.8$ at the beginning of the cooling track, to $7.2$ at the end of core crystallization (see their Figure~1). These compositions are vastly different than the mass fraction ratio of $^{12}$C/$^{16}$O $=$ $1$ that is commonly assumed in studies of CO novae.

We do not know the actual $^{12}$C/$^{16}$O mass fraction ratio in the outer white dwarf core that gave rise to the isotopic signatures in a given nova presolar grain. Therefore, we will randomly sample this parameter over the range predicted by white dwarf models ($1.5$ $\leq$ $^{12}$C/$^{16}$O $\leq$ $8.0$) to see which values, if any, reproduce the stardust grain data. 

Another problem with assuming an outer core composition of equal $^{12}$C and $^{16}$O abundances is that classical novae are expected to recur on time scales of $\approx$ $10^4$ $-$ $10^5$~yr. Each nova outburst leaves behind a remnant layer composed of helium and other nuclear burning products, and of unburned hydrogen, on top of the outer white dwarf core (light grey area in Figure~\ref{fig:sketch}a and c). Since this layer takes part in the burning during the next flash \citep{fujimoto92,prialnik95}, it impacts the composition of the nuclear fuel. 

\subsubsection{Mixing between white dwarf and accreted matter during the flash}
Spectroscopic observations show that CNO elements are considerably enriched relative to hydrogen in many nova ejecta \citep{gehrz98}. This enrichment plays a critical role for the dynamic ejection of a portion of the envelope and presumably results from mixing of the outer core white dwarf matter with the accreted matter (red region in Figure~\ref{fig:sketch}b). 
Recent two- and three-dimensional simulations now yield encouraging results by demonstrating that Kelvin-Helmholtz instabilities during the thermonuclear runaway can lead to an enrichment of the accreted envelope with material from the underlying white dwarf at levels that approximately agree with observations \citep{glasner95,glasner12,casanova11,casanova16}. 

So far, multi-dimensional nova simulations incorporate very small nuclear reaction networks ($\approx$ $30$ nuclides only), and are not suitable for studying the nucleosynthesis in detail. Consequently, one-dimensional simulations are indispensible for this purpose, but cannot account self-consistently for the mixing at the interface between the outer white dwarf core and the accreted matter. Most one-dimensional simulations work around this problem by artificially enriching the envelope with outer core matter to a predetermined degree. The enriched matter is then accreted and its history is followed through the nuclear burning and mass ejection stages.

Frequently, one-dimensional nova simulations assume that the accreted matter from the companion and the white dwarf core material pre-mix with equal mass fractions. Recent work, albeit in the context of ONe novae, hinted at a pre-mixing fraction of $25$\% white dwarf matter and $75$\% accreted matter \citep{kelly13}, which provides a significantly better fit to the measured elemental abundances in several, but not all, nova ejecta. 

While these are encouraging first steps, we are far from being able to predict the degree of element mixing in a given observed nova. In particular, the pre-mixing parameter, $f_{pre}$, is poorly constrained at present. We define it by mixing one part of outer white dwarf core matter with $f_{pre}$ parts of accreted solar-like matter, 
\begin{equation}
X_{pre} \equiv \frac{X_{WD} + f_{pre} X_{acc}}{1+f_{pre}}
\label{eq:premix}
\end{equation}
where $X_i$ denotes mass fraction. We will randomly sample this parameter over a reasonable range ($0.5$ $\leq$ $f_{pre}$ $\leq$ $5.0$) to see which values best reproduce the stardust grain data. The outer bounds of this interval correspond to white dwarf admixtures of 67\% and 17\%, respectively.

\subsubsection{Dilution of the ejecta}\label{sec:postmix}
We already mentioned that previous classical nova simulations result in much more anomalous isotopic ratios compared to the values measured in nova candidate grains. To explain the observations, an additional mixing episode, after the outburst, has been postulated (which we will term ``post-mixing''), whereby the ejecta processed by nuclear burning mix with more than ten times the amount of unprocessed, solar-like matter before grain condensation \citep{amari01}. However, the source and mechanism of this potential dilution is not understood. 

The grains we are considering here condensed at temperatures well over $1000$~K and, therefore, it is difficult to explain the formation of SiC or olivine grains by mixing of nova ejecta with the interstellar medium. 

One idea was pursued by \citet{figueira17}, who studied the collision of the nova ejecta initially with the accretion disk and subsequently with the companion. They found that the matter escaping from the binary system is predominantly composed of the ejecta, i.e., the contribution of the accretion desk or the companion is small. Under these conditions, we expect on average only minor post-mixing, although enhanced dilution may perhaps occur in local regions. If grains can condense in this environment, we nevertheless expect that only a small fraction will show signatures of significant post-mixing. 

More studies of this important issue are needed, since we neither know the source of the solar-like matter for post-mixing, nor if any post-mixing took place at all. At this time, we conclude the following. If we consider two measured stardust grains, and for one grain all data agree with CO nova simulations without the need for any post-mixing, while the other grain requires dilution to match data to the CO nova simulations, then the former grain is more likely to originate from a CO nova. We will return to this argument in Section~\ref{sec:allresults}.


Since the post-mixing process is poorly constrained, our simulations will account for it using a post-mixing parameter, $f_{post}$, defined by
\begin{equation}
X_{post} \equiv \frac{X_{proc} + f_{post} X_{pris}}{1+f_{post}}
\label{eq:postmix}
\end{equation}
where $X_{proc}$ and $X_{pris}$ denote the mass fractions from the reaction network output (i.e., processed matter) and the pristine matter (i.e., solar-like), respectively. We will sample this parameter over a range of $0$ $\leq$ $f_{post}$ $\leq$ $10^4$ to see which values best reproduce the stardust grain data.

\subsection{Comparison of simulations to presolar stardust grain data}
Isotopic data for all presolar stardust grains that have been suggested over the years to originate from novae are compiled in Table~\ref{tab:graindata}. The data are separated according to grain chemistry (SiC, silicate, graphite, and oxide). As already noted above, we have no unambiguous evidence linking any of these grains to a nova paternity. Neither can we exclude unambiguously a nova paternity for many other grains among the thousands of stardust samples measured so far. Our goal is to investigate the conditions, if any, that could give rise to the measured isotopic anomalies.

The values listed for C, N, and O represent isotopic number abundance ratios, while for Mg, Si, and S the data correspond to parts-per-thousand deviations from solar matter, e.g.,
\begin{equation}
\begin{split}
& \delta \left( ^{25}\mathrm{Mg}/^{24}\mathrm{Mg} \right) \equiv \delta ^{25}\mathrm{Mg}  \\ 
& \equiv\left[   \frac{ \left( ^{25}\mathrm{Mg}/^{24}\mathrm{Mg} \right)_{exp} }{\left( ^{25}\mathrm{Mg}/^{24}\mathrm{Mg} \right)_\odot}   -1 \right] \times 1000
\end{split}
\label{eq:delta}
\end{equation}
where the most abundant isotope of the element (i.e., of even mass number) appears in the denominator. 

We chose to exclude consideration of $\delta^{26}\mathrm{Mg}$ values and inferred $^{26}$Al/$^{27}$Al ratios. First, for grains with significant amounts of both Mg and Al, some of the observed $^{26}$Mg may have condensed originally as $^{26}$Al. In such cases, we cannot simply add the simulated $^{26}$Mg and $^{26}$Al abundances and compare the total to the measured $\delta^{26}\mathrm{Mg}$ values because the Al and Mg may have condensed in a particular grain with different efficiencies. Although, in principle, this can be taken into account if the Al/Mg ratio of a grain is known, e.g., for MgAl$_2$O$_4$ grains \citep{zinner05}, this ratio has not been reported for many of the grains. Second, even for grains like SiC with high Al/Mg ratios and hence purely radiogenic $^{26}$Mg, Al contamination has been shown to have impacted nearly all reported isotopic measurements \citep{zinner13,groopman15}, leading to an underestimate of the original $^{26}$Al/$^{27}$Al ratios by up to a factor of two. The magnitude of such contamination can be estimated for a given grain through detailed analysis of the original measurement data \citep{groopman15}, but this information has not been reported for the nova candidate grains of this study.


When comparing grain measurements to results from nucleosynthesis simulations, two important issues need to be addressed. First, we cannot reasonably expect that a simulation will precisely reproduce the grain measurements, since there are too many approximations involved in any nucleosynthesis model. If the simulation results are ``close'' to the data, say, within some factor, we may accept the computed results as a possible solution. Second, we need to account for the systematic bias in the grain measurements, in addition to the statistical uncertainty that is included in the reported error. Systematic effects arise from contamination, e.g., from sampling the meteorite material surrounding the grain, or from sample preparation. It is important to emphasize that this bias could move a data point into the direction of less anomalous values only, i.e., contamination will not make a grain appear more anomalous than it really is. 

We will adopt the following procedure for determining approximate agreement between simulation and measurement (``acceptable solutions''). The simulated mass fraction of a nuclide with atomic number $Z$ and mass number $A$, denoted by $X_{sim}(^AZ)$, is divided and multiplied by a systematic uncertainty factor, $n_{sim}$, according to
\begin{equation}\label{eq:delta4}
\begin{split}
X_{sim}(^AZ)/n_{sim} \leq X_{sim}(^AZ) \\
\leq X_{sim}(^AZ) \times n_{sim} 
\end{split}
\end{equation}
This range is then transformed into a range of simulated $\delta$-values, $\delta^AZ_{sim}$, 
\begin{equation}\label{eq:delta2}
[ \delta^AZ_{sim}^{low} , \delta^AZ_{sim}^{high} ]
\end{equation}
Next, using an experimental uncertainty factor, $n_{exp}$, we define a range for a measured value of $\delta^{A}Z_{exp}^{mean} \pm \delta^{A}Z_{exp}^{err}$,
\begin{equation}\label{eq:delta3}
\begin{split}
[ n_{exp}^{(1-\pi)/2} \times \delta^{A}Z_{exp}^{mean} - n_{exp} \times \delta^{A}Z_{exp}^{err},\\
n_{exp}^{(1+\pi)/2} \times \delta^{A}Z_{exp}^{mean} + n_{exp} \times \delta^{A}Z_{exp}^{err} ]
\end{split}
\end{equation}
where $\pi$ $=$ $sign(\delta^{A}Z_{exp}^{mean})$ = $\pm1$ denotes the sign of the experimental mean value. We define an acceptable solution if the two regions given by Equations~\ref{eq:delta2} and \ref{eq:delta3} overlap. For the two factors containing the effects of the simulation and measurement bias, we adopt values of $n_{sim}$ $=$ $1.7$ and $n_{exp}$ $=$ $2$. The former value implies an uncertainty factor for a simulated abundance {\it ratio} of $1.7^2$ $\approx$ $3$. It was chosen to exceed the factor of $2$ within which our one-zone simulations reproduce the results of the multi-zone hydrodynamic calculation (Section~\ref{sec:scheme}). As already pointed out in Section~\ref{sec:strategy}, we will only accept solutions for which simulated and measured $\delta$-values overlap for {\it all measured isotopic ratios}. 
 
To gain a better understanding of this procedure, consider the following numerical example. Suppose a value of $\delta^{13}C_{exp}^{mean} \pm \delta^{13}C_{exp}^{err}$ $=$ $12734 \pm 167$ has been measured for a hypothetical grain. Accounting both for statistical and systematic effects, we translate this experimental result into an experimental interval of $[12400,25802]$, according to Equation~\ref{eq:delta3}. Furthermore, suppose that one among many reaction network runs results in final mass fractions of $X_{sim}(^{12}C)$ $=$ $0.203$ and $X_{sim}(^{13}C)$ $=$ $0.0193$. According to Equation~\ref{eq:delta4}, these values are translated to ranges of $0.1194$ $\leq$ $X_{sim}(^{12}C)$ $\leq$ $0.3451$ and $0.011353$ $\leq$ $X_{sim}(^{13}C)$ $\leq$ $0.03281$. The interval for the corresponding $X_{sim}(^{13}C)/X_{sim}(^{12}C)$ ratios is then given by $[ 0.03289, 0.2748 ]$, resulting in a range of simulated $\delta^{13}C_{sim}$ values of $ [ 1701, 21568 ] $. In this case, the experimental and simulation ranges overlap. Only if the same applies to all other isotopic ratios measured for this particular grain do we retain the solutions (i.e., the model parameters, such as peak temperatures and densities, mixing parameters, etc.) of this particular reaction network calculation.
 
\section{Results}\label{sec:results}
We will first show what results can be obtained with our method by using grain LAP-149 as an example. We then summarize results for all nova candidate grains. Finally we discuss those grains that most likely originate from CO novae.

\subsection{Example: grain LAP-149}\label{sec:lap149}
The nova candidate graphite grain LAP-149 has a diameter of about $1$~$\mu m$ and exhibits one of the lowest $^{12}$C/$^{13}$C ratios ever measured (Table~\ref{tab:graindata}). The $^{14}$N/$^{15}$N ratio is high, but the oxygen, silicon, and sulfur isotopic ratios are close to solar within experimental uncertainties. \citet{haenecour16} found that the C, N, Si, and S isotopic ratios could be reproduced by a CO nova model involving a $0.6$~$M_\odot$ white dwarf with 50\% pre-mixing ($f_{pre}$ $=$ $1$; Equation~\ref{eq:premix}) without assuming any post-mixing. However, the measured and simulated $^{17}$O/$^{16}$O and $^{18}$O/$^{16}$O ratios disagreed by orders of magnitude. The other CO nova models, for white dwarf masses in the range of $M_{WD}$ $=$ $0.8$ $-$ $1.15$~$M_\odot$, did not provide a match to any of the measured isotopic ratios. Notice that LAP-149 is the only nova candidate grain with eight measured isotopic ratios (see Table~\ref{tab:graindata}).

\begin{figure*}
\epsscale{2.0}
\plotone{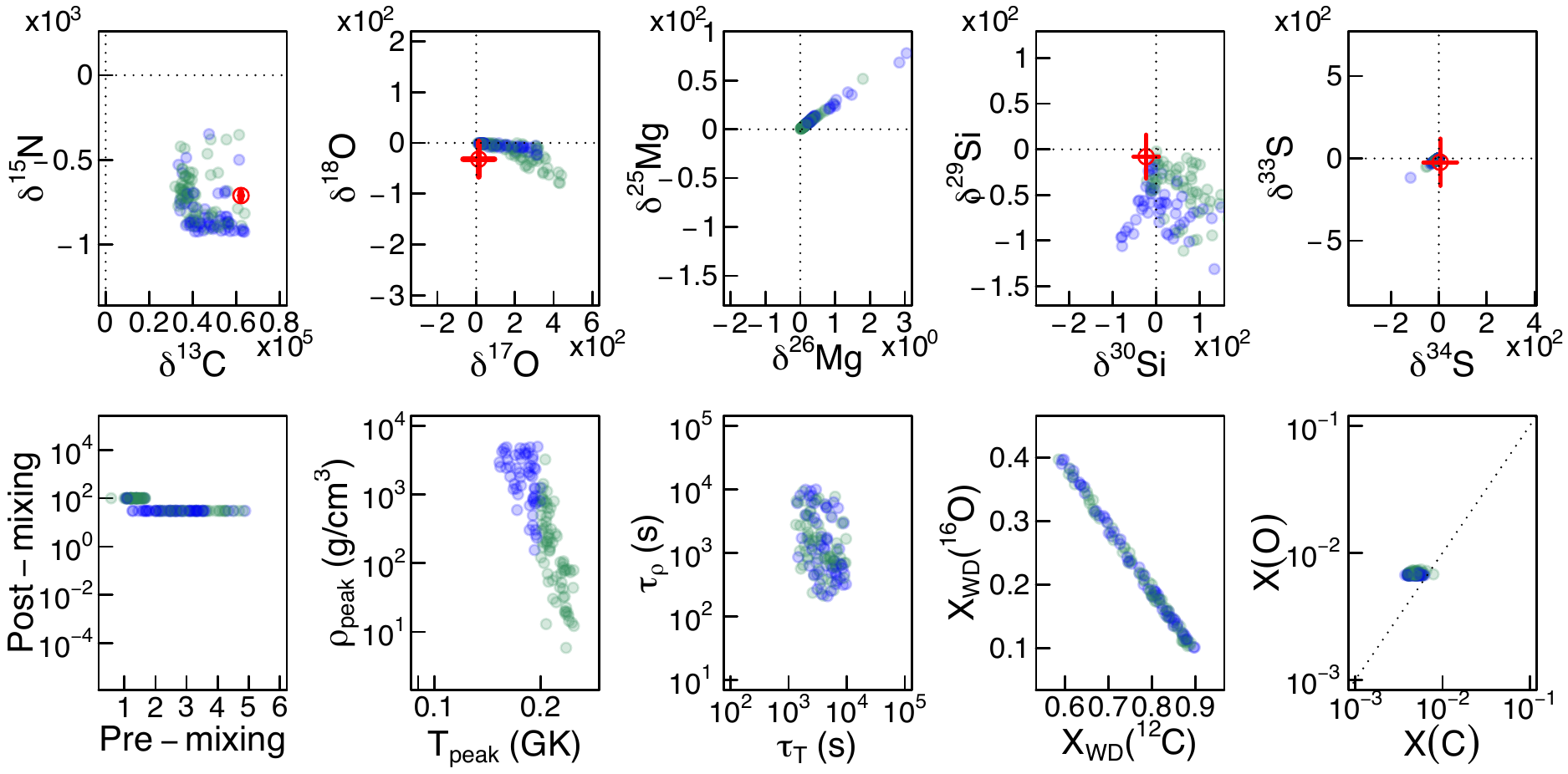}
\caption{Summary results for the presolar stardust graphite grain LAP-149, the only nova candidate grain with eight different measured isotopic ratios. (Top row) Isotopic ratios for C, N, O, Mg, Si, and S. The data points are shown in red; simulations that reproduce all data simultaneously (see Equations~\ref{eq:delta2} and \ref{eq:delta3}) are displayed in blue and green, depending on the sampled peak temperature (see below). (Bottom row) From left to right: post-mixing fraction versus pre-mixing; peak density versus peak temperature; 1/e exponential decay times for the density evolution versus temperature evolution; outer white dwarf mass fractions of $^{16}$O versus $^{12}$C; and elemental oxygen versus carbon mass fractions {\it after} post-mixing. The blue and green simulation results correspond to peak temperatures of T$_{peak}$ $<$ $0.20$~GK and T$_{peak}$ $\ge$ $0.20$~GK, respectively. The simulation results shown were obtained with recommended nuclear interaction rates only, i.e., without Monte Carlo sampling of the nuclear rates.
\label{fig:lap149}}
\end{figure*}

Our results for LAP-149 are shown in Figure~\ref{fig:lap149}, which was obtained after computing 25,000 network samples. The top row displays the measured isotopic ratios (red) together with the simulations. Only those simulation results are displayed that simultaneously agree with all data, according to Equations~\ref{eq:delta2} and \ref{eq:delta3}. The different colors for the simulation results correspond to different peak temperature values (blue: T$_{peak}$ $<$ $0.20$~GK, green: T$_{peak}$ $\ge$ $0.20$~GK). The corresponding sampled model parameters are shown in the bottom row.

We obtain acceptable solutions for a wide range of pre-mixing fractions, between $f_{pre}$ $=$ $1$ and $f_{pre}$ $=$ $5$ (first bottom panel), corresponding to outer white dwarf core admixtures between 50\% and 16\%, respectively. Post-mixing fractions are in the range of $f_{post}$ $=$ $30$ $-$ $100$, implying a significant admixture of solar-like matter after the explosion. In particular, no acceptable solutions are obtained without post-mixing, in agreement with the findings of \citet{haenecour16}. Acceptable peak temperature and peak density values (second bottom panel) scatter throughout the sampled ranges ($150$~MK $\le$ T$_{peak}$ $\le$ $250$~MK, $5$~g/cm$^3$ $\le$ $\rho_{peak}$ $\le$ $5\times10^3$~g/cm$^3$). The 1/e exponential decay times for temperature and density (third bottom panel) scatter within the range of $10^3$~s $\le$ $\tau_T$ $\le$ $10^4$~s and $10^2$~s $\le$ $\tau_\rho$ $\le$ $10^4$~s, respectively. Solutions are obtained for the entire range of sampled outer white dwarf core composition, $0.6$ $\le$ X$_{WD}(^{12}C)$ $\le$ $0.9$, $0.1$ $\le$ X$_{WD}(^{16}O)$ $\le$ $0.4$ (fourth bottom panel). Lastly, the ratios of elemental carbon to oxygen, after post-mixing, are in the range of X(C)/X(O) $=$ $0.6$ $-$ $1.0$ (fifth bottom panel).

So far, we used in the simulations for all rates of thermonuclear reactions and weak interactions the recommended (i.e., median) values provided by STARLIB. However, the nuclear rates have uncertainties, either derived from experimental nuclear physics input or from theoretical models (Section~\ref{sec:rates}). For this reason, we repeated the above Monte Carlo procedure of computing 25,000 network samples, but this time included the random sampling of the nuclear rates according to their probability densities contained in STARLIB (Section~\ref{sec:procedure}). As a result, the scatter of the simulation points (black, blue, green) in Figure~\ref{fig:lap149} increased slightly, but all relevant features discussed above are preserved. In other words, current reaction rate uncertainties have only a small impact on our results for this particular grain.

The above results do not {\it prove} a CO nova origin for LAP-149, because we are not considering here the production in sites other than novae (e.g., supernovae or AGB stars). But we can conclude that this grain could only have been produced by the temperature and density conditions, and compositions, typical of CO novae if the ejecta mixed with a large fraction of solar-like matter, i.e., $1$ part of ejecta with at least $30$ parts of solar-like matter. 

We carefully repeated the random sampling using different random number seeds and total numbers of simulations. All results shown in this work, including Figure~\ref{fig:lap149}, are robust, in the sense that they are reproducible apart from small statistical fluctuations.

\subsection{Results for all nova candidate grains}\label{sec:allresults}
Similar to grain LAP-149 discussed in the previous section, we investigated for each of the $39$ grains listed in Table~\ref{tab:graindata} if our simulations can reproduce the measured isotopic ratios according to Equations~\ref{eq:delta2} and \ref{eq:delta3}. Three points need to be considered for the following discussion. 

First, the number of measured isotopic ratios will strongly impact the likelihood that a given grain originates from a CO nova. In other words, if simulations reproduce to a similar degree the data for grains A and B, and only two isotopic ratios have been measured in grain A compared to six ratios in grain B, then the latter grain is more likely to be of CO nova paternity. Clearly, the more isotopic ratios measured, the tighter the constraints on the grain origin. 

Second, the relative number of network runs (``acceptable solutions'') that provide simultaneous solutions for all measured isotopic ratios of a given grain is also important in this regard. We cannot claim that the {\it absolute} number of acceptable solutions reflects the probability of a CO nova paternity. But we can conclude that a low number of solutions indicates a fine-tuning of model parameters, while a high number of solutions results from model parameter combinations that occupy a larger volume of the parameter space. In other words, the {\it relative} number of acceptable solutions reflects the likelihood of a CO nova paternity.

Third, we discussed in Section~\ref{sec:postmix} that the origin and mechanism for a possible dilution of the ejecta by solar-like matter (``post-mixing'') is not well understood, and that it is likely that a fraction of CO nova grains condense in the ejecta with little, if any, post-mixing. For this reason, we assume that a given grain has a higher chance of a CO nova paternity if its measured isotopic ratios can be simulated without any post-mixing.

If we allow for post-mixing of various degrees, we find acceptable solutions for almost all grains listed in Table~\ref{tab:graindata}. The only exceptions are grains 8-9-3 and KFC1a-551. For these, not a single acceptable solution is obtained, and thus a CO nova paternity is highly unlikely. Also, it would be a fortuitous coincidence if all the other $37$ stardust grains listed in Table~\ref{tab:graindata} would be of CO nova origin. Therefore, we conclude that only a weak case can be made for a CO nova paternity if significant post-mixing must be invoked to match observed and simulated isotopic ratios. 

Table~\ref{tab:paternity} lists all grains for which we found acceptable solutions {\it without assuming any post-mixing.} For each grain we show the mineralogy, the number of measured isotopic ratios, and the measured elements. The last column shows the total number of network runs that provide simultaneous solutions for all measured isotopic ratios of a given grain according to Equations~\ref{eq:delta2} and \ref{eq:delta3}. The grains are rank ordered, from top to bottom, according to the plausibility of a CO nova paternity. As discussed above, we ranked the grains according to the number of measured isotopic ratios (column 3) and the number of acceptable simulations (column 5).

We find that six grains, all of them of the SiC variety, have a high plausibility of a CO nova origin (from  G270-2 to Ag2$\_$6 in Table~\ref{tab:paternity}). These grains will be discussed in more detail in Section~\ref{sec:highplaus}. In this case, we obtain acceptable solutions without any post-mixing, and we have several measured isotopic ratios ($\geq$4) or many acceptable solutions ($\geq$10).

The next group consists of twelve grains (from Ag2 to 1$\_$07 in Table~\ref{tab:paternity}) that do not require any post-mixing. These have a medium plausibility of a CO nova paternity. We rank them below the top group because they either have a small number of measured isotopic ratios (i.e., fewer experimental constraints) or a small number of acceptable simulations. 

\begin{figure*}
\epsscale{2.0}
\plotone{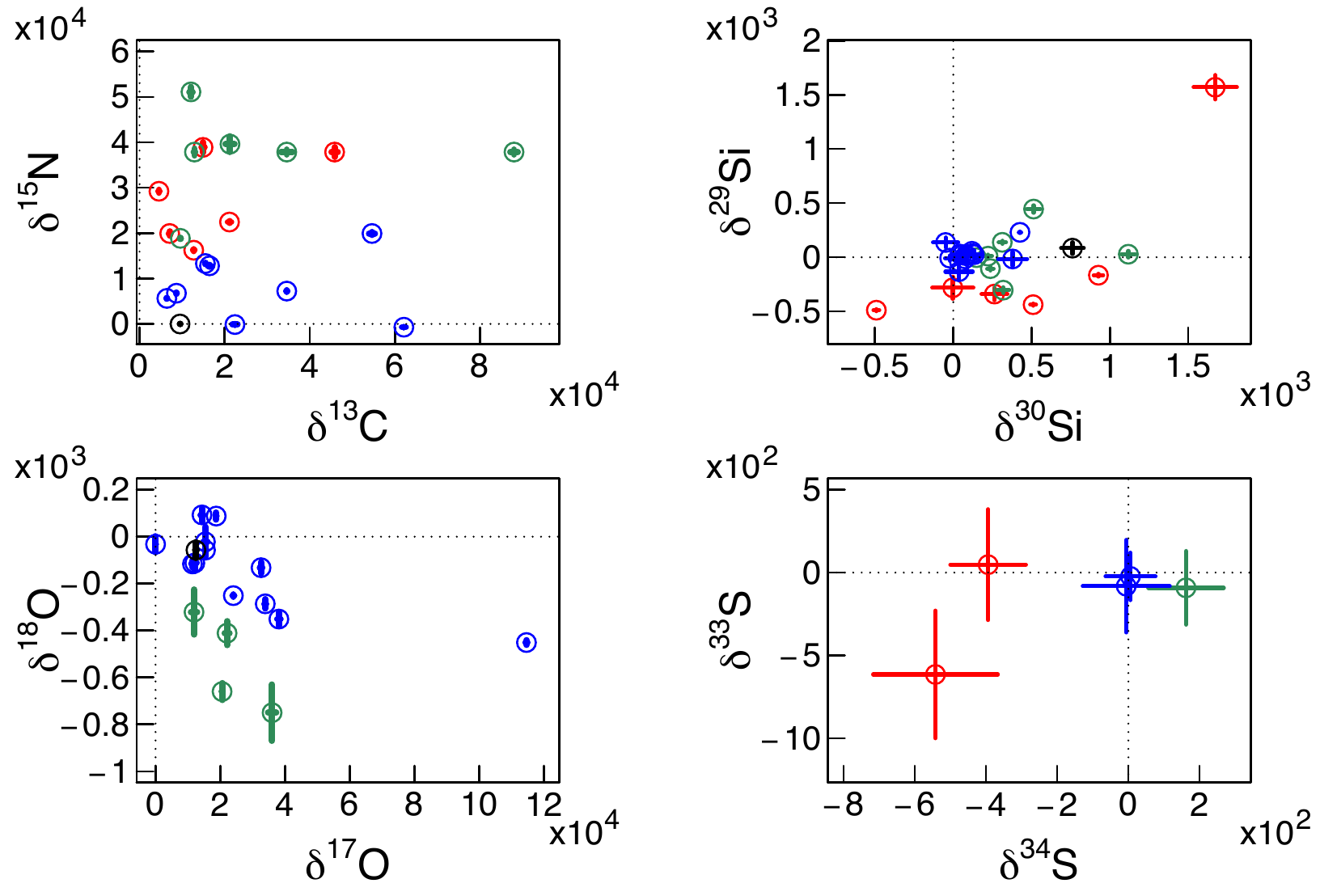}
\caption{Measured isotopic ratios for the elements C, N, Si, and S (see Table~\ref{tab:graindata}). The colors red and green indicate grains of high and medium plausibility, respectively, of a CO nova paternity (see discussion in the text); for these two groups, acceptable solutions are obtained without requiring any post-mixing of the ejecta. Grains shown in blue require post-mixing to match the measured isotopic ratios and correspond to a low plausibility of a CO nova paternity. For the two grains shown in black, no solutions are obtained with or without post-mixing of the ejecta; thus it is highly unlikely that they originate from CO novae.
\label{fig:allgrains}}
\end{figure*}

Figure~\ref{fig:allgrains} shows the measured C, N, O, Si, and S isotopic ratios of all grains listed in Table~\ref{tab:graindata}. The colors red and green indicate grains of high and medium plausibility, respectively, of a CO nova paternity. For these two groups, acceptable solutions are obtained without assuming any post-mixing of the ejecta. Grains shown in blue require post-mixing to match the measured isotopic ratios and correspond to a low plausibility of a CO nova paternity. For the two grains shown in black, no solutions are obtained with or without post-mixing of the ejecta, and thus they most likely do not originate from CO novae.

\subsection{High-plausibility CO nova grains}\label{sec:highplaus}
The SiC grains G270-2, G278, Ag2\_6 \citep{liu16}, and M11-334-2, M11-347-4, M11-151-4 \citep{nittler05}, shown in boldface in Table~\ref{tab:paternity}, have the highest plausibility of a CO nova paternity. For these grains, between four and six isotopic ratios of the elements C, N, Si, and S have been measured, and our simulations sampling the CO nova parameter space provide simultaneous solutions to all data without requiring any dilution of the ejecta with solar-like matter. 

Interestingly, most of these have been argued to originate from supernovae rather than novae on the basis of their isotopic signatures. For example, M11-334-2 has $^{28}$Si, $^{44}$Ca, and $^{49}$Ti excesses and a very high inferred initial $^{26}$Al abundance, similar to those seen in type X SiC grains from supernovae, and M11-151-4 has an unexplained $^{47}$Ti anomaly. The $^{32}$S excesses seen in G270-2 and AG2\_6 and the strong $^{28}$Si depletion seen in G278 have not been predicted by previous nova models.

\begin{figure*}
\epsscale{2.0}
\plotone{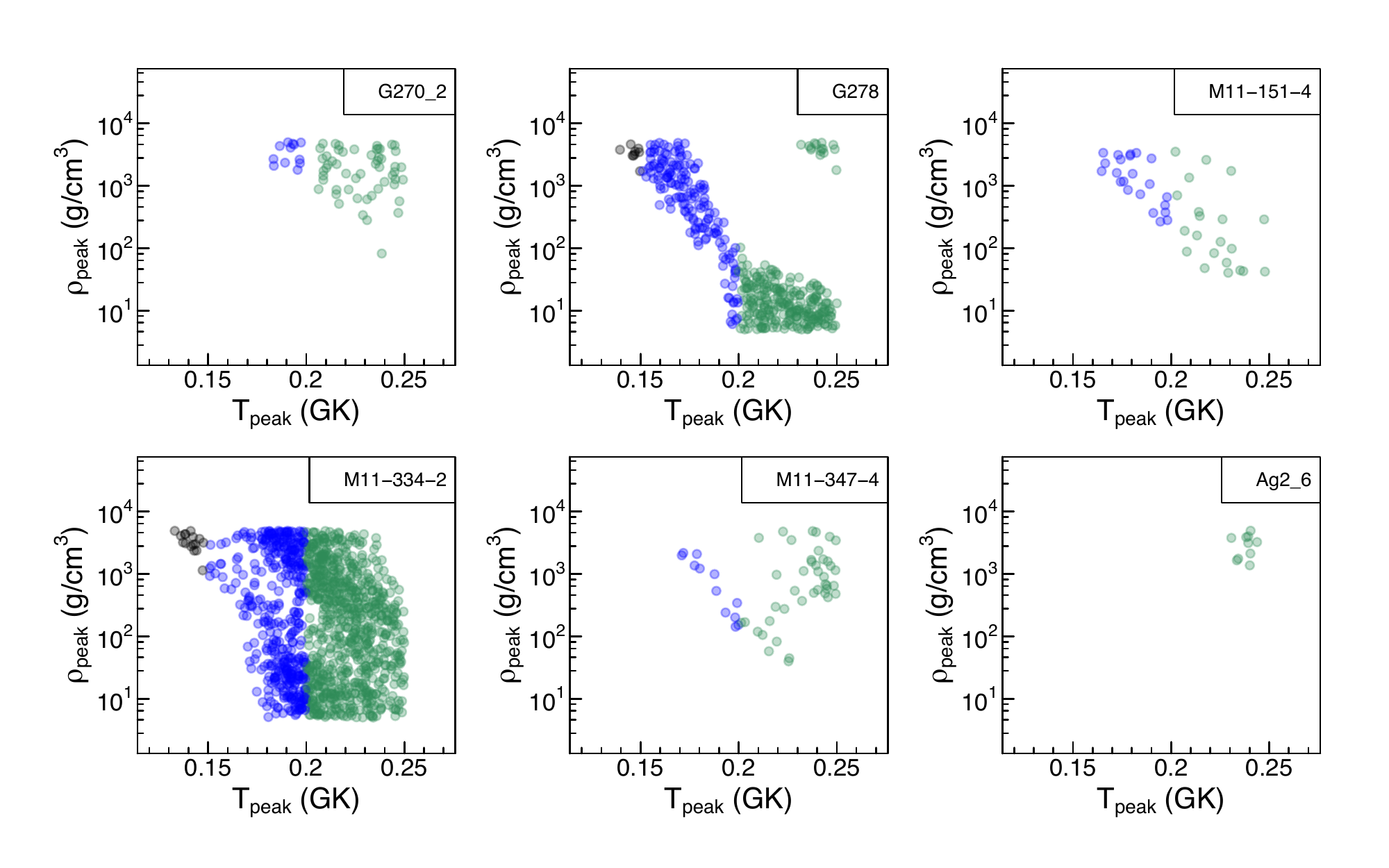}
\caption{Peak density, $\rho_{peak}$, versus peak temperature, T$_{peak}$, for the six presolar stardust grains with the highest plausibility of a CO nova paternity. These are shown in boldface in column 5 of Table~\ref{tab:paternity}. The colors have the same meaning as in Figure~\ref{fig:lap149}. The simulation results, using 50,000 network calculations, were obtained assuming no post-mixing of ejecta with solar-like matter ($f_{post}$ $=$ $0$) and without any variations of thermonuclear reaction rates.
\label{fig:highgrains1}}
\end{figure*}

\begin{figure*}
\epsscale{2.0}
\plotone{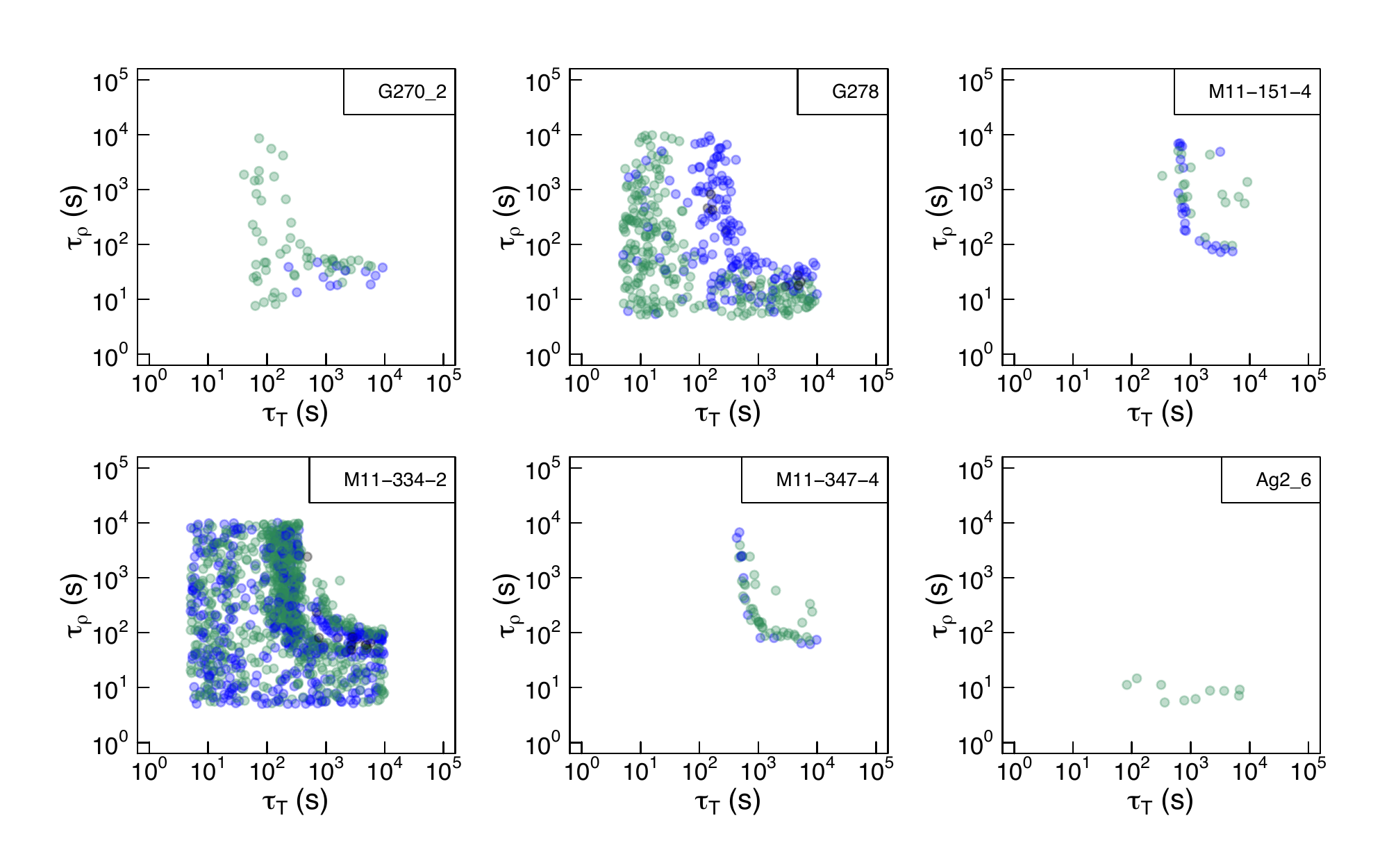}
\caption{Exponential 1/e decay time scale of the density profile versus the exponential 1/e decay time scale of the temperature profile, for the six presolar stardust grains with the highest plausibility of a CO nova paternity (shown in boldface in column 5 of Table~\ref{tab:paternity}). The colors have the same meaning as in Figure~\ref{fig:lap149}. The simulation results, using 50,000 network calculations, were obtained assuming no post-mixing of ejecta with solar-like matter ($f_{post}$ $=$ $0$) and without any variations of thermonuclear reaction rates.
\label{fig:highgrains2}}
\end{figure*}

We will now consider the simulated CO nova peak temperature and peak density conditions that are obtained for these grains, assuming no post-mixing. They are shown in Figure~\ref{fig:highgrains1}, using the same color scheme that was employed in Figure~\ref{fig:lap149} (black, blue, and green for T$_{peak}$ $\le$ $0.15$~GK, $0.15$~GK $<$ T$_{peak}$ $<$ $0.20$~GK, and T$_{peak}$ $\ge$ $0.20$~GK, respectively). The simulation results are not uniformly spread over the T$_{peak}$ $-$ $\rho_{peak}$ plane. Instead, the solutions occupy select regions. Black simulation points are not apparent, except for a small number of points for grains G278 and M11-334-2. This indicates that all six grains likely originate in nova explosions with peak temperatures in excess of $150$~MK, involving higher-mass CO white dwarfs. For grains G270$\_$2 and Ag2$\_$6, the most likely peak temperature exceeds $200$~MK, as can be seen from the relative number of green simulation points. Figure~\ref{fig:highgrains2} shows for the same six grains the exponential 1/e decay time scale of the density profile versus the exponential 1/e decay time scale of the temperature profile. The simulation points occupy a parameter space typical for CO nova conditions. 

\begin{figure*}
\epsscale{2.0}
\plotone{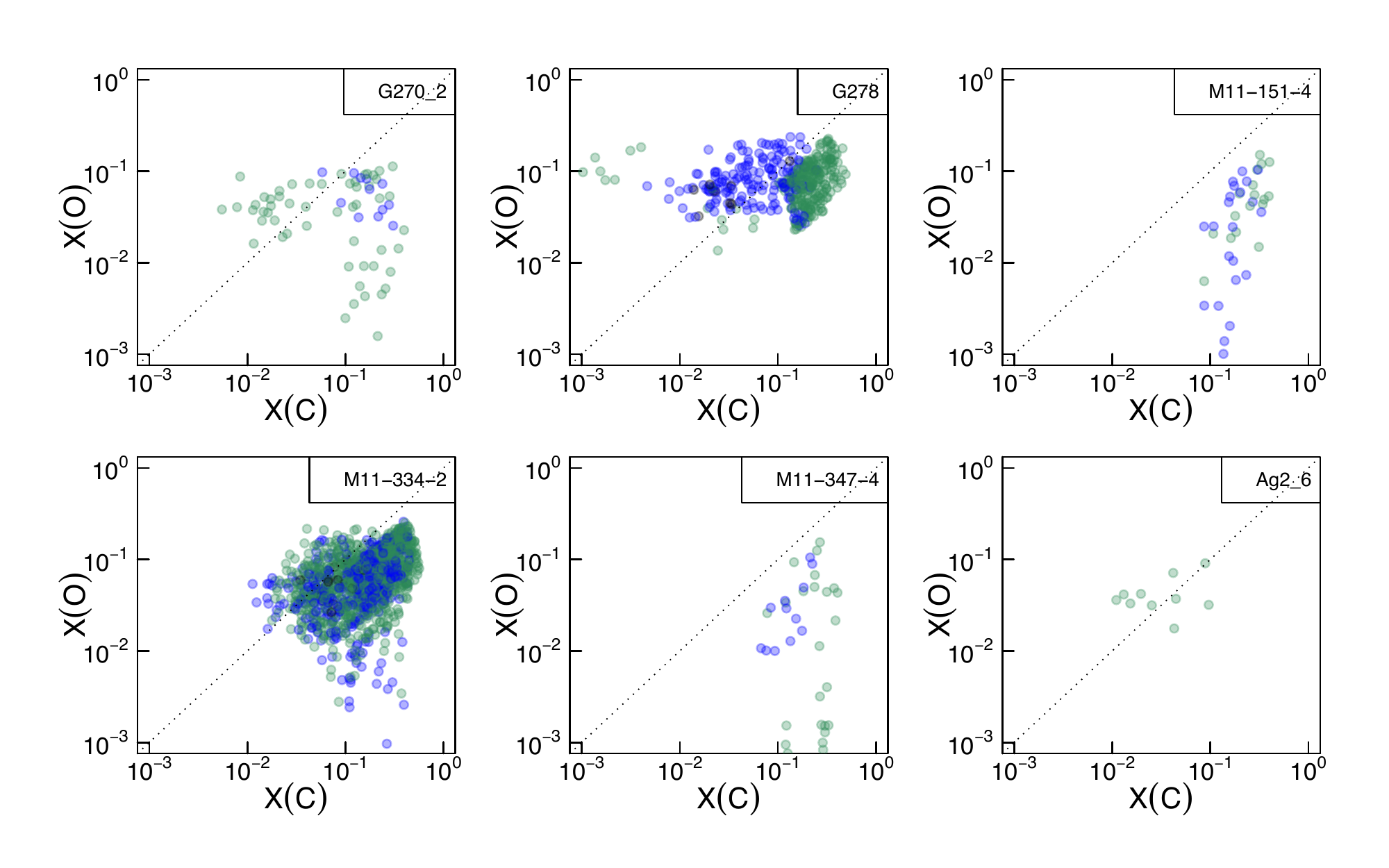}
\caption{Elemental oxygen versus elemental carbon abundance (by mass) in the ejecta, for the six presolar stardust grains with the highest plausibility of a CO nova paternity (shown in boldface in column 5 of Table~\ref{tab:paternity}). The colors have the same meaning as in Figure~\ref{fig:lap149}. The simulation results, using 50,000 network calculations, were obtained assuming no post-mixing of ejecta with solar-like matter ($f_{post}$ $=$ $0$) and without any variations of thermonuclear reaction rates. The dotted lines correspond to equal mass fractions.
\label{fig:highgrains3}}
\end{figure*}

It is interesting to consider the measured elemental carbon and oxygen abundances in CO nova ejecta, and compare the observations to the simulations. The observational results are summarized in Table~\ref{tab:dust}. The carbon-to-oxygen mass fraction ratios range from $0.084$ (for GQ Mus) to $5.4$ (for V827 Her). Dust has been directly observed in PW Vul, QV Vul, V827 Her, V842 Cen, and V1668 Cyg (column 5). At least two of these, QV Vul and V842 Cen, have produced SiC dust. The simulated elemental oxygen versus  carbon abundances (by mass) are shown in Figure~\ref{fig:highgrains3}, without assuming any post-mixing. Most of the simulation results scatter about the dotted line, corresponding to equal carbon and oxygen mass fractions. A few simulation points, for grain G278 only, exhibit ratios of X(O)/X(C)$\lesssim$ $0.3$ (i.e., the points on the far left in the second top panel), which would be unfavorable for the condensation of SiC grains. For all solutions shown in Figure~\ref{fig:highgrains3}, the median of the elemental silicon mass fraction amounts to X$_{med}$(Si) $\approx$ $6\times10^{-4}$, which is close to the solar value, X$_\odot$(Si) $\approx$ $8\times10^{-4}$.

\begin{figure*}
\epsscale{2.0}
\plotone{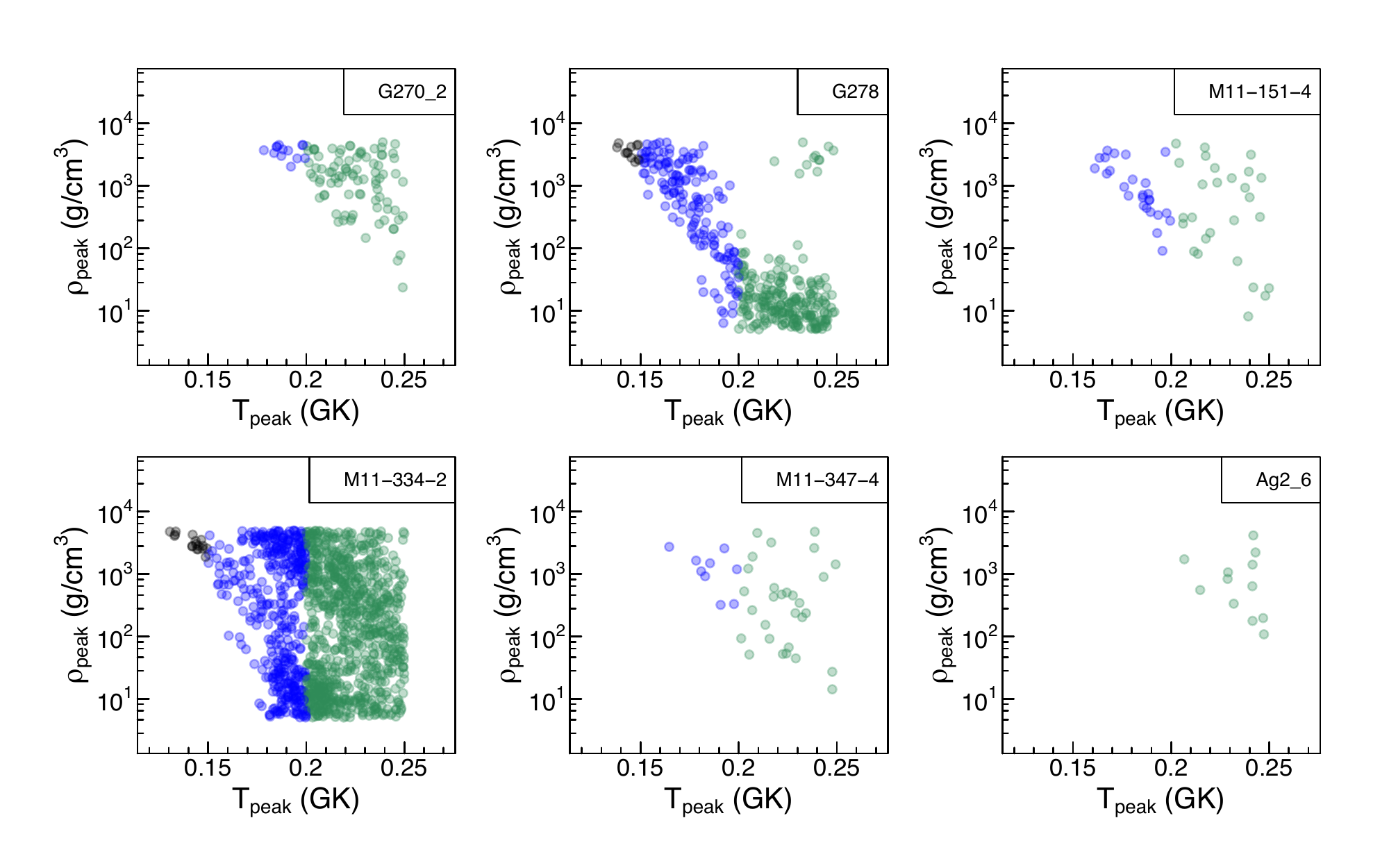}
\caption{Same as Figure~\ref{fig:highgrains1}, except that the random sampling of thermonuclear reaction rates is included in the results. An increase in the scatter of the simulation points (black, blue, green) compared to Figure~\ref{fig:highgrains1} is apparent.
\label{fig:highgrains4}}
\end{figure*}

Finally, we repeated the simulations under exactly the same conditions, except that we included this time thermonuclear reaction rate variations. As mentioned in Section~\ref{sec:rates}, all reaction rates in the network were sampled simultaneously according to their rate probability densities given by STARLIB. The results for the peak density versus peak temperature are shown in Figure~\ref{fig:highgrains4}. Comparison to Figure~\ref{fig:highgrains1}, which was obtained without thermonuclear rate variations, shows that the thermonuclear rate uncertainties increase the scatter of the simulation points (black, blue, green) noticeably. A detailed analysis of which nuclear reaction rate variations have the largest impact on the scatter is beyond the scope of the present work and will be presented in a forthcoming publication. Nevertheless, all relevant features discussed above are preserved, even when taking into account thermonuclear rate variations.







\section{Summary}\label{sec:summary}
We discussed a new method to analyze the CO nova paternity of presolar stardust grains. Previously, a small number of  simulations was performed, assuming fixed values for key parameters (e.g., white dwarf composition, mixing during the explosion, peak temperatures and densities, explosion time scales, dilution of the ejecta after the outburst, and thermonuclear reaction rates) that are sometimes weakly constrained. Such a method will always result in a poor exploration of the large nova parameter space and, consequently, in a weak predictive power of the applied model. In addition, previous comparisons between predicted and observed isotopic abundance ratios were not based on a rigorous statistical procedure, resulting in conflicting claims of CO nova paternities. 

In this work, we applied a Monte Carlo method by randomly sampling over realistic parameter ranges. We adopted thermonuclear reaction rates, and their associated probability densities, from the STARLIB library. We also provide a statistical interpretation for what we mean by ``agreement'' or ``disagreement'' between observed and predicted isotopic ratios (Equations~\ref{eq:delta2} and \ref{eq:delta3}).

Based on the numerical results for the parameters of our model, we identify $18$ presolar grains with measured isotopic signatures consistent with a CO nova origin, {\it without requiring any dilution of the ejecta}. Among these, the grains G270$\_$2, M11-334-2, G278, M11-347-4, M11-151-4, and Ag2$\_$6 have the highest plausibility of a CO nova paternity. As is the case with all previous studies, we cannot determine the {\it absolute} probability that any given grain originates in a CO nova. Such a conclusion can only be drawn if a study similar to the one presented here is applied to competing astrophysical scenarios, e.g., supernovae and AGB stars. Such an investigation is planned for the future. Numerical results for any of the grains listed in Table~\ref{tab:graindata} can be requested from the first author.

\acknowledgments
We would like to thank Andrea Derdzinski, Bob Gehrz,  Ann Nguyen, David Little, Jack Dermigny, and Maitrayee Bose for helpful comments. This work was supported by the U.S. Department of Energy under Contract No. DE-FG02-97ER41041 and by NASA under the Astrophysics Theory Program Grant 14-ATP14-0007. JJ ackowledges  partial support by the Spanish MINECO through grant AYA2014-59084-P, and by the AGAUR/Generalitat de Catalunya grant SGR0038/2014.

\clearpage

\begin{deluxetable}{lccccc}
\tabletypesize{\scriptsize}
\tablecaption{Carbon, oxygen, and dust observations in CO novae.\tablenotemark{a}\label{tab:dust}}
\tablewidth{0pt}
\tablehead{
\colhead{CO nova}           & \colhead{X$_{obs}$(C)\tablenotemark{b}}        & \colhead{X$_{obs}$(O)\tablenotemark{b}}       & \colhead{X$_{obs}$(C)/X$_{obs}$(O)}       & \colhead{Type of dust\tablenotemark{c}}  &  Mass of dust (M$_\odot$)
}
\startdata
GQ Mus		& 0.0080				& 0.095				& 0.084				& none detected				& \nodata	\\
HR Del		& \nodata				& 0.047				& \nodata				& \nodata						& \nodata	\\
LMC 1991		& \nodata				& \nodata				& 0.27\tablenotemark{f}	& \nodata						& \nodata	\\
LW Ser		& \nodata				& \nodata				& \nodata				& C							& 3.6$\times10^{-7}$ \\
NQ Vul		& \nodata				& \nodata				& \nodata				& C							& 2$\times10^{-7}$ \\
PW Vul		& 0.031				& 0.047				& 0.66				& C							& 5.1$\times10^{-10}$	\\
QV Vul		& \nodata				& 0.041				& \nodata				& C, SiO$_2$, HC, SiC			& 1.0$\times10^{-6}$\tablenotemark{j}	\\
V339 Del		& \nodata				& \nodata				& \nodata				& \nodata						&  5$\times10^{-9}$\tablenotemark{m}	\\
V443 Sct		& \nodata				& 0.007				& \nodata				& \nodata						& \nodata	\\
V705 Cas\tablenotemark{n}	& \nodata	& \nodata				& \nodata				& C, HC, SiO$_2$				& 8.2$\times10^{-7}$\tablenotemark{j}		\\
V827 Her		& 0.087				& 0.016				& 5.4					& C							& \nodata	\\
V842 Cen		& 0.12				& 0.03				& 4.0					& C, SiC, HC					& \nodata	\\
V1186 Sco	& \nodata				& \nodata				& \nodata				& none detected\tablenotemark{i}	& \nodata	\\
V1280 Sco	&  \nodata				&  \nodata				&  \nodata				&  C, SiO$_2$\tablenotemark{g}	&  1.0$\times10^{-7}$\tablenotemark{l}    \\
V1425 Aql		& 0.030\tablenotemark{d}	& 0.085\tablenotemark{d}	& 0.35				& \nodata						& \nodata	\\
V1668 Cyg	& \nodata				& \nodata				& \nodata				& C							& 2.1$\times10^{-8}$\tablenotemark{j}	\\
V2214 Oph	& \nodata				& 0.060				& \nodata				& \nodata						& \nodata	\\
V2362 Cyg	& \nodata				& 0.163\tablenotemark{e}	& \nodata				& \nodata						& $\approx2\times10^{-10}$ $-$ $2\times10^{-8}$\tablenotemark{k} 	\\  
V2676 Oph	& \nodata				& \nodata				& \nodata				& C, SiO$_2$\tablenotemark{h}	 	& \nodata	\\
\enddata
\tablenotetext{a}{From \citet{gehrz98}, unless noted otherwise; if more than one value is quoted, we adopt the arithmetic average value. Only those novae are listed for which the white dwarf paternity (i.e., CO) has been established.} 
\tablenotetext{b}{Abundance by mass.} 
\tablenotetext{c}{C=amorphous carbon; HC=hydrocarbons; SiO$_2$=silicate.} 
\tablenotetext{d}{From \citet{lyke01}.} 
\tablenotetext{e}{From \citet{munari08}.}
\tablenotetext{f}{From \citet{schwarz01}.}
\tablenotetext{g}{From \citet{sakon16}.}
\tablenotetext{h}{From \citet{kawakita17}; also reported by \citet{kawakita15}: $^{12}$C/$^{13}$C $\approx$ $4$ and $^{14}$N/$^{15}$N $\approx$ $2$.}
\tablenotetext{i}{From \citet{schwarz07}.}
\tablenotetext{j}{From \citet{gehrz08}.}
\tablenotetext{k}{From \citet{arai10}.}
\tablenotetext{l}{From \citet{chesneau08}.}
\tablenotetext{m}{From \citet{evans17}.}
\tablenotetext{n}{\citet{hric98} suggest a white dwarf mass of $0.79\pm0.06$~M$_\odot$.}

\end{deluxetable}

\clearpage

\newgeometry{margin=1.5cm}
\begin{deluxetable}{lccccccccc}  
\tabletypesize{\tiny}
\tablecolumns{10}
\tablewidth{0pt}
\rotate
\tablecaption{Measured isotopic ratios in nova candidate presolar stardust grains.\label{tab:graindata}}
\renewcommand\tabcolsep{2pt}
\tablehead{   
  \colhead{\multirow{2}{*}{Grain}} &
  \colhead{\multirow{2}{*}{$^{12}$C/$^{13}$C}} &
  \colhead{\multirow{2}{*}{$^{14}$N/$^{15}$N}} &
  \colhead{$^{17}$O/$^{16}$O} &
  \colhead{$^{18}$O/$^{16}$O} &
  \colhead{\multirow{2}{*}{$\delta^{25}$Mg/$^{24}$Mg}} &
  \colhead{\multirow{2}{*}{$\delta^{29}$Si/$^{28}$Si}} &
  \colhead{\multirow{2}{*}{$\delta^{30}$Si/$^{28}$Si}} &  
  \colhead{\multirow{2}{*}{$\delta^{33}$S/$^{32}$S}} &
  \colhead{\multirow{2}{*}{$\delta^{34}$S/$^{32}$S}} \\
  \colhead{\phantom} &
  \colhead{\phantom} &
  \colhead{\phantom} &
  \colhead{($\times$10$^{-4}$)} &
  \colhead{($\times$10$^{-3}$)} &
  \colhead{\phantom} &
  \colhead{\phantom} &
  \colhead{\phantom} &
} 
\startdata
\multicolumn{10}{c}{SiC Grains}\\
\hline
AF15bB-429-3$^{1}$		& 9.4 $\pm$ 0.2	& \nodata			& \nodata			& \nodata			& \nodata			& 28 $\pm$ 30			& 1118 $\pm$ 44		& \nodata			& \nodata\\
AF15bC-126-3$^{1}$	& 6.8 $\pm$ 0.2	& 5.22 $\pm$ 0.11	& \nodata			& \nodata			& \nodata			& -105 $\pm$ 17		& 237 $\pm$ 20		& \nodata			& \nodata\\
Ag2$^{2}$				& 2.5 $\pm$ 0.1	& 7.0 $\pm$ 0.1	& \nodata			& \nodata			& \nodata			& -304 $\pm$ 26		& 319 $\pm$ 38		& -92 $\pm$ 222	& 162 $\pm$ 106\\ 
Ag2\_6$^{2}$			& 16.0 $\pm$ 0.4	& 9.0 $\pm$ 0.1	& \nodata			& \nodata			& \nodata			& -340 $\pm$ 57		& 263 $\pm$ 82		& 48 $\pm$ 334	& -394 $\pm$ 106\\
KJC112$^{1}$			& 4.0 $\pm$ 0.2	& 6.7 $\pm$ 0.3	& \nodata			& \nodata			& \nodata			& \nodata				& \nodata				& \nodata			& \nodata\\
KJGM4C-100-3$^{1}$	& 5.1 $\pm$ 0.1	& 19.7 $\pm$ 0.3	& \nodata			& \nodata			& \nodata			& 55 $\pm$ 5			& 119 $\pm$ 6			& \nodata			& \nodata\\
KJGM4C-311-6$^{1}$	& 8.4 $\pm$ 0.1	& 13.7 $\pm$ 0.1	& \nodata			& \nodata			& \nodata			& -4 $\pm$ 5			& 149 $\pm$ 6			& \nodata			& \nodata\\
G1614$^{2}$			& 9.2 $\pm$ 0.07	& 35.0 $\pm$ 0.7	& \nodata			& \nodata			& \nodata			& 34 $\pm$ 5			& 121 $\pm$ 6			& \nodata			& \nodata\\
G1697$^{2}$			& 2.5 $\pm$ 0.01	& 33.0 $\pm$ 0.8	& \nodata			& \nodata			& \nodata			& -42 $\pm$ 12			& 40 $\pm$ 15			& \nodata			& \nodata\\
G1748$^{2}$			& 5.4 $\pm$ 0.02	& 19.0 $\pm$ 0.2	& \nodata			& \nodata			& \nodata			& 21 $\pm$ 4			& 83 $\pm$ 5			& \nodata			& \nodata\\
G270\_2$^{2}$			& 11.0 $\pm$ 0.3	& 13.0 $\pm$ 0.3	& \nodata			& \nodata			& \nodata			& -282 $\pm$ 101		& -3 $\pm$ 131			& -615 $\pm$ 385	& -542 $\pm$ 175\\
G283$^{2}$			& 12.0 $\pm$ 0.1	& 41.0 $\pm$ 0.5	& \nodata			& \nodata			& \nodata			& -15 $\pm$ 3			& 75 $\pm$ 4			& \nodata			& \nodata\\
G278$^{2}$			& 1.90 $\pm$ 0.03	& 7.0 $\pm$ 0.2	& \nodata			& \nodata			& \nodata			& 1570 $\pm$ 112		& 1673 $\pm$ 138		& \nodata			& \nodata\\
G1342$^{2}$			& 6.40 $\pm$ 0.08	& 7.00 $\pm$ 0.14	& \nodata			& \nodata			& \nodata			& 445 $\pm$ 34		& 513 $\pm$ 43		& \nodata			& \nodata\\
GAB$^{2}$			& 1.60 $\pm$ 0.02	& 13.0 $\pm$ 0.2	& \nodata			& \nodata			& \nodata			& 230 $\pm$ 6			& 426 $\pm$ 7			& -82 $\pm$ 279	& -6 $\pm$ 122\\
G240-1$^{2}$			& 1.00 $\pm$ 0.01	& 7.0 $\pm$ 0.1	& \nodata			& \nodata			& \nodata			& 138 $\pm$ 14		& 313 $\pm$ 23		& \nodata			& \nodata\\
M11-151-4$^{3,}$\tablenotemark{a}		& 4.02 $\pm$ 0.07	& 11.6 $\pm$ 0.1	& \nodata			& \nodata			& \nodata			& -438 $\pm$ 9			& 510 $\pm$ 18		& \nodata			& \nodata\\
M11-334-2$^{3,}$\tablenotemark{b}		& 6.48 $\pm$ 0.08	& 15.8 $\pm$ 0.2	& \nodata			& \nodata			& \nodata			& -489 $\pm$ 9			& -491 $\pm$ 18		& \nodata			& \nodata\\
M11-347-4$^{3}$		& 5.59 $\pm$ 0.13	& 6.8 $\pm$ 0.2	& \nodata			& \nodata			& \nodata			& -166 $\pm$ 12		& 927 $\pm$ 30		& \nodata			& \nodata\\
M26a-53-8$^{4}$		&  4.75 $\pm$ 0.23	& \nodata			& \nodata			& \nodata			& \nodata			& 10 $\pm$ 13			& 222 $\pm$ 25		& \nodata			& \nodata\\
\hline
\multicolumn{10}{c}{Silicate Grains}\\
\hline
1\_07$^{5}$			& \nodata			& \nodata			& 49.1 $\pm$ 3.6	& 1.36 $\pm$ 0.19	& \nodata			& \nodata				& \nodata				& \nodata			& \nodata\\
4\_2$^{6}$			& \nodata			& \nodata			& 128.0 $\pm$ 1.4	& 1.74 $\pm$ 0.05	& 1025 $\pm$ 29	& 24 $\pm$ 40			& 134 $\pm$ 52		& \nodata			& \nodata\\
4\_7$^{6}$			& \nodata			& \nodata			& 149.0 $\pm$ 2.0	& 1.30 $\pm$ 0.06	& 213 $\pm$ 56	& 136 $\pm$ 46		& -49 $\pm$ 80			& \nodata			& \nodata\\
A094\_TS6$^{7}$		& \nodata			& \nodata			& 95.4 $\pm$ 1.1	& 1.50 $\pm$ 0.01	& \nodata			& 29 $\pm$ 43			& 43 $\pm$ 54			& \nodata			& \nodata\\
AH-106a$^{8}$			& \nodata			& \nodata			& 50.1 $\pm$ 2.2	& 1.78 $\pm$ 0.07	& \nodata			& 15 $\pm$ 59			& 80 $\pm$ 67			& \nodata			& \nodata\\
B2-7$^{9}$			& \nodata			& \nodata			& 133.0 $\pm$ 1.0	& 1.43 $\pm$ 0.04	& \nodata			& 21 $\pm$ 56			& 57 $\pm$ 69			& \nodata			& \nodata\\
GR95\_13\_29$^{10}$	& \nodata			& \nodata			& 62.5 $\pm$ 2.5	& 1.96 $\pm$ 0.14	& 79 $\pm$ 21		& -16 $\pm$ 63			& 379 $\pm$ 92		& \nodata			& \nodata\\
\hline
\multicolumn{10}{c}{Graphite Grains}\\
\hline
KFB1a-161$^{4,11}$		& 3.8 $\pm$ 0.1	& 312 $\pm$ 43	& \nodata			& \nodata			& -28 $\pm$ 62 	& -133 $\pm$ 81		& 37 $\pm$ 87			& \nodata			& \nodata\\
KFC1a-551$^{4,}$\tablenotemark{c}		& 8.46 $\pm$ 0.04	& 273 $\pm$ 8		& \nodata			& \nodata			& -157 $\pm$ 443	& 84 $\pm$ 54		& 761 $\pm$ 72		& \nodata			& \nodata\\
LAP-149$^{12}$		& 1.41 $\pm$ 0.01	& 941 $\pm$ 81	& 3.86 $\pm$ 0.34	& 1.94 $\pm$ 0.07	& \nodata			& -8 $\pm$ 24			& -23 $\pm$ 29			& -23 $\pm$ 143	& 6 $\pm$ 70\\	
\hline
\multicolumn{10}{c}{Oxide Grains}\\
\hline
12-20-10$^{13}$		& \nodata			& \nodata			& 88.0 $\pm$ 3.0	& 1.18 $\pm$ 0.11	& \nodata			& \nodata				& \nodata				& \nodata			& \nodata\\
8-9-3$^{13}$			& \nodata			& \nodata			& 51.4 $\pm$ 1.1	& 1.89 $\pm$ 0.07	& -66 $\pm$ 21		& \nodata				& \nodata				& \nodata			& \nodata\\
C4-8$^{13}$			& \nodata			& \nodata			& 440.4 $\pm$1.2	& 1.10 $\pm$ 0.02	& 949 $\pm$ 8		& \nodata				& \nodata				& \nodata			& \nodata\\
KC23$^{14}$			& \nodata			& \nodata			& 58.5 $\pm$ 1.8	& 2.19 $\pm$ 0.06	& 45 $\pm$ 35		& \nodata				& \nodata				& \nodata			& \nodata\\
KC33$^{14}$			& \nodata			& \nodata			& 82.2 $\pm$ 0.6	& 0.68 $\pm$ 0.08	& \nodata			& \nodata				& \nodata				& \nodata			& \nodata\\
MCG67$^{10}$			& \nodata			& \nodata			& 47.3 $\pm$ 1.4	& 1.77 $\pm$ 0.03	& \nodata			& \nodata				& \nodata				& \nodata			& \nodata\\
MCG68$^{10}$			& \nodata			& \nodata			& 62.6 $\pm$1.1	& 1.89 $\pm$ 0.02	& \nodata			& \nodata				& \nodata				& \nodata			& \nodata\\
S-C6087$^{15}$		& \nodata			& \nodata			& 75.2 $\pm$ 0.3	& 2.18 $\pm$ 0.03	& 36 $\pm$ 22		& \nodata				& \nodata				& \nodata			& \nodata\\
T54$^{16}$			& \nodata			& \nodata			& 141 $\pm$ 5		& 0.5 $\pm$ 0.2	& \nodata			& \nodata				& \nodata				& \nodata			& \nodata\\
\hline
Solar					& 89				& 272			& 3.8				& 2.0				& 0.0				& 0.0					& 0.0					& 0.0				& 0.0\\ 
\hline
\enddata
\tablecomments{Presented errors are 1$\sigma$. Some ratios are presented as deviations from solar abundances in permil, $\delta^{i}$X/$^{j}$X $\equiv$ [($^{i}$X/$^{j}$X)/($^{i}$X/$^{j}$X)$_{\odot}$--1]$\times$1000. $^{12}$C/$^{13}$C and $^{14}$N/$^{15}$N (ratio in air) are from \citet{jose04} and references therein; $^{17}$O/$^{16}$O and $^{18}$O/$^{16}$O solar values are from \citet{leitner12}.}

\tablenotetext{a}{Additional isotopic ratios for Grain M11-151-4 are given in \citet{nittler05}: $^{26}$Al/$^{27}$Al = 0.27 $\pm$ 0.05, $\delta^{46}$Ti/$\delta^{48}$Ti = 28 $\pm$ 59, $\delta^{47}$Ti/$\delta^{48}$Ti = 215 $\pm$ 57, $\delta^{49}$Ti/$\delta^{48}$Ti = 82 $\pm$ 55, $\delta^{50}$Ti/$\delta^{48}$Ti = -100 $\pm$ 123.}

\tablenotetext{b}{Additional isotopic ratios for Grain M11-334-2 are given in \citet{nittler05}: $^{26}$Al/$^{27}$Al = 0.39 $\pm$ 0.06, $\delta^{42}$Ca/$\delta^{40}$Ca = -70 $\pm$ 200, $\delta^{44}$Ca/$\delta^{40}$Ca = 535 $\pm$ 150, $\delta^{46}$Ti/$\delta^{48}$Ti = -61 $\pm$ 33, $\delta^{47}$Ti/$\delta^{48}$Ti = -5 $\pm$ 36, $\delta^{49}$Ti/$\delta^{48}$Ti = 380 $\pm$ 47, $\delta^{50}$Ti/$\delta^{48}$Ti = -20 $\pm$ 59.}

\tablenotetext{c}{Multiple values exist for the $^{16}$O/$^{18}$O (or $^{18}$O/$^{16}$O) ratio of Grain KFC1a-551 and are not given here (see: \citet{amari95}; M. Bose, priv. comm. (2017); and http://presolar.wustl.edu/\~{} pgd).}

\tablerefs{$^{1}$\citet{amari01}; $^{2}$\citet{liu16}; $^{3}$\citet{nittler05}; $^{4}$\citet{nittler03}; $^{5}$\citet{vollmer07}; $^{6}$\citet{nguyen14}; $^{7}$\citet{nguyen07}; $^{8}$\citet{nguyen10}; $^{9}$\citet{bose10}; $^{10}$\citet{leitner12}; $^{11}$\citet{jose07}; $^{12}$\citet{haenecour16}; $^{13}$\citet{gyngard10}; $^{14}$\citet{nittler08}; $^{15}$\citet{choi99}; $^{16}$\citet{nittler97}.
}
\end{deluxetable}

\clearpage
\restoregeometry

\begin{deluxetable}{lcccr}
\tabletypesize{\scriptsize}
\tablecaption{Summary results of our simulations\tablenotemark{a}. The order, from top to bottom, reflects approximately the likelihood that a given presolar stardust grain originated from a CO nova.\label{tab:paternity}}
\tablewidth{0pt}
\tablehead{
\colhead{Grain\tablenotemark{b}}           & \colhead{Mineralogy\tablenotemark{b}}        & \colhead{Number of}       & \colhead{Measured}                                
                        &       \colhead{Number of}       \\
                                    &                                         & \colhead{isotopic ratios\tablenotemark{c}} 
                                    &  \colhead{elements\tablenotemark{c}}        &    \colhead{solutions\tablenotemark{d}}     
}
\startdata
{\bf G270-2}              & SiC        &   6      &   C, N, Si, S         &	 67		                  \\
{\bf M11-334-2}         & SiC        &   4      &  C, N, Si               &    1228	                 \\
{\bf G278}			& SiC         &   4      &  C, N, Si		     	&	 425		   \\
{\bf M11-347-4}         & SiC        &   4      &  C, N, Si               &     56		                \\
{\bf M11-151-4}         & SiC        &   4      &  C, N, Si              &    43 		                \\
{\bf Ag2\_6} 		& SiC 	  &   6     &   C, N, Si, S     	   & 	 10		                   \\
Ag2                   	& SiC         &   6     &   C, N, Si, S         	&    3		           \\
G1342			& SiC	  &    4     &   C, N, Si		     &	 11		   \\
AF15bC-126-3  	& SiC         &   4     &   C, N, Si             &    5   		                  \\
G240-1			& SiC	  &    4     &   C, N, Si		     &	 8		   \\
KJGM4C-311-6 	& SiC         &   4     &  C, N, Si               &	 2		          \\
AF15bB-429-3  	& SiC         &   3     &   C, Si                   &    102		    \\
M26a-53-8        	& SiC         &   3      &  C, Si                   &    7		    \\
T54                    	& oxide      &   2     &  O                         &   11048	   \\
KC33                 	& oxide      &  2      &  O                         &   8840	    \\
KJC112             	& SiC         &    2    &  C, N                    &    1315 	     \\
12\_20\_10        	& oxide      &   2     &  O                         &   330		   \\
1\_07                 	& silicate   &  2       &  O                         &   3		    \\
\enddata
\tablenotetext{a}{The table shows all presolar stardust grains for which we found acceptable solutions without assuming any post-mixing, i.e., without any dilution of the ejecta before grain condensation.} 
\tablenotetext{b}{See Table~\ref{tab:graindata}; the grains in boldface have the highest plausibility for a CO nova paternity.} 
\tablenotetext{c}{Measured number of isotopic ratios and elements in grain.} 
\tablenotetext{d}{The number of network runs, out of a total of 50,000 simulations, that provide simultaneous solutions for all measured isotopic ratios of a given grain.} 
\end{deluxetable}

\bibliographystyle{apalike}
\bibliography{tableref}
\clearpage

\end{document}